\let\myover=\over
\documentstyle[a4,epsfig,11pt,axodraw,amssymb]{article}
\newcommand{\smU}{{\scriptscriptstyle U}}
\newcommand{\sm}[1]{{\scriptscriptstyle #1}}

\def\l{\left(}
\def\r{\right)}
\renewcommand{\Im}{\mathop{\rm Im}\nolimits}
\renewcommand{\Re}{\mathop{\rm Re}\nolimits}
\let\over=\myover
\begin{document}

%%%%%%%%%%%%% Title
\begin{center}
  {\Large\bf Flavor violation and $\tan{\beta}$ in gauge mediated models with 
messenger-matter mixing} \\
  \medskip
  S.~L.~Dubovsky\footnote{E-mail: \verb|sergd@ms2.inr.ac.ru|},
  D.~S.~Gorbunov\footnote{E-mail: \verb|gorby@ms2.inr.ac.ru|}\\
  \medskip
  {\small
     Institute for Nuclear Research of
         the Russian Academy of Sciences, 117312 Moscow, 
Russia
  }
\end{center}

\begin{abstract}
  
We consider the Minimal Gauge Mediated Model (MGMM) with either
fundamental or antisymmetric messenger multiplets and study
consequences of mes\-sen\-ger-matter mixing. We find that in these
models, unlike MGMM without mixing, wide range of $\tan\beta$ is
allowed.  It is shown that existing experimental limits on processes
with lepton flavor violation, and on $K^0-\bar{K}^0$, $D^0-\bar{D}^0$ and
$B^0-\bar{B}^0$ - mixing place significant constraints on relevant
coupling constants and mixing parameters. On the other hand, the
contributions of the messenger-matter mixing to the rates of $\tau \to
e\gamma$, $\tau
\to \mu\gamma$ and $b\to s\gamma$ may be well below the present 
experimental limits depending on the value of $\tan\beta$. 
We also point out the possibility of sizeable
slepton oscillations in this model.
\end{abstract}

\section{Introduction}
Presently, much attention is paid to flavor physics in supersymmetric
theories.  Flavor violation naturally emerges in those SUSY models
where supersymmetry is broken by supergravity interactions. The
corresponding soft-breaking terms are often assumed to be universal at
the Planck (string) scale. This universality breaks down due to the
renormalization group evolution between the Planck (string) and GUT
scales~\cite{Universality_violation}.  As a result, sizeable mixing
in slepton and squark sectors at low energies is induced.  This mixing leads to
lepton flavor violation for ordinary leptons ($\mu \to e\gamma$, etc.)
and FCNC processes ($K^0-\bar{K}^0$-mixing, etc.) at rates close to
existing experimental limits \cite{barbieri}.

Another class of SUSY models invokes gauge mediated supersymmetry
breaking \cite{dine}. In these models, gauge interactions do not
lead to flavor violation because the messenger-matter interactions are
flavor blind.  Nevertheless, there is a way to introduce flavor
changing interactions in these theories. This possibility is based on
the observation that some of the messenger fields have the same
quantum numbers as some of the Standard Model (SM) fields. Therefore, it is
natural to consider direct mixing between messenger and matter fields
\cite{dinem}.  In such variants of the gauge mediated models,
messengers not only transfer SUSY breaking to usual matter, but also
generate flavor violation.  Another effect of this mixing is the
absence of heavy stable charged (and colored) particles (messengers)
in the theory~\cite{dinem}.  Other known possibilities to solve the latter
problem~\cite{Dimopoulos} faced the necessity of 
fine-tuning of the messenger mass parameters.

The purpose of this paper is to show that in a reasonable range of
parameters, messenger-matter mixing in the Minimal Gauge Mediated Model
(MGMM)\cite{kolda} can give rise to the observable rates of rare
lepton flavor violating processes like $\mu~\to~e\gamma$ and $\mu - e$
conversion and can play a significant role in quark flavor physics.
Also we study the effect of messenger-matter mixing on radiative
electroweak symmetry breaking. 
We will consider the messengers belonging either to fundamental or to 
antisymmetric
tensor representation of the $SU(5)$ GUT group.

We will see that the tree level mixing between the Standard Model
fermions is small due to see-saw type suppression. The tree level
mixing between sparticles is also small in the part of the parameter
space which is natural for MGMM. However, we observe that radiative
corrections involving interactions with ordinary Higgs sector induce
much stronger mixing of the scalars of MSSM. The point is that
messengers obtain masses not through interactions with ordinary Higgs
fields, but through interactions with hidden sector. So, the overall
matrices of trilinear couplings of sleptons (squarks) and messengers
with Higgs fields are not proportional to the corresponding mass
matrices (unlike in the Standard Model). As a consequence, in the
basis of eigenstates of the tree level mass matrix, matrices of
trilinear couplings are not diagonal in flavor.  The largest trilinear
terms involve squarks and sleptons as well as messenger and Higgs
superfields.  It is these terms that cause slepton and additional
squark mixing\footnote{These mixing terms are additional to the
ordinary mixing in the squark sector of MSSM appearing due to the
Yukawa couplings to the Higgs fields.} at the one loop level through
loops with Higgs fields and messengers.

We find out that, unlike MGMM without mixing (where $\tan{\beta}$ is
large, $\tan{\beta}\gtrsim 50$~\cite{kolda}), there is a wide region
of allowed $\tan{\beta}$ in the model with mixing. The value of
$\tan{\beta}$ is determined by the magnitude of the mixing terms. This
observation gives rise to an interesting possibility to relate the rates of
flavor violating processes to the Higgs sector parameters.
 
At high $\tan{\beta}$, messenger masses of order $10^5$~GeV and
messenger-matter Yukawa couplings of order $10^{-1}-10^{-3}$, the
rates of $\mu~\to~e\gamma$ and $\mu - e$ conversion and flavor
violating $\tau$ decays are comparable to their experimental
limits. Similar result applies to additional FCNC processes in quark
sector. At low $\tan{\beta}$ the experimental constraints on the
mixing terms are weaker by about an order of magnitude.  It turns out
that theoretical constraints inherent in the model imply that at low
$\tan{\beta}$ the contribution of messenger-matter mixing to $b\to
s\gamma$ process is negligible and $\tau\to e\gamma$ and
$\tau\to\mu\gamma$ decay rates are much below the present experimental
limits. The latter observation means that the discovery of lepton
flavor violation in $\tau$ decays will rule out models with
messenger-matter mixing and low $\tan{\beta}$.

In Ref.~\cite{we}, the lepton mixing was explored in the model with
messengers in the fundamental representation of $SU(5)$. In fact, the
analysis of Ref.~\cite{we} is valid at low
$\tan{\beta}$. Here we study radiative electroweak breaking and extend
the previous analysis to the quark sector and antisymmetric messenger
representation as well as to high $\tan{\beta}$.

It is
worth mentioning that the constraints coming from the flavor changing
processes, as presented here, may be used for qualitative estimates of
the mixing parameters not only in MGMM with mixing,
but also in more general gauge mediated models.

\section{The model}
\label{s2}
\begin{figure}[tb]

\begin{picture}(0,0)%
\epsfig{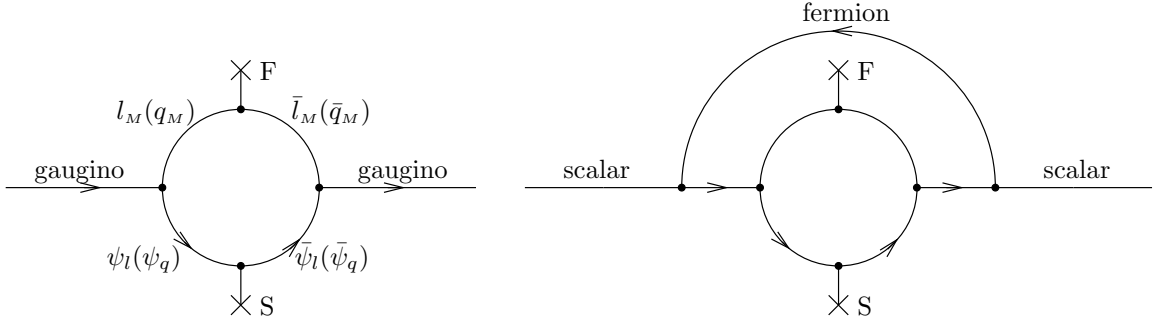}%
\end{picture}%
\setlength{\unitlength}{3947sp}%
\begingroup\makeatletter\ifx\SetFigFont\undefined%
\gdef\SetFigFont#1#2#3#4#5{%
  \reset@font\fontsize{#1}{#2pt}%
  \fontfamily{#3}\fontseries{#4}\fontshape{#5}%
  \selectfont}%
\fi\endgroup%
\begin{picture}(7224,1956)(589,-1527)
\put(2200,-28){\makebox(0,0)[lb]{\smash{\SetFigFont{10}{12.0}{\familydefault}{\mddefault}{\updefault}F}}}
\put(2200,-1505){\makebox(0,0)[lb]{\smash{\SetFigFont{10}{12.0}{\familydefault}{\mddefault}{\updefault}S}}}
\put(5955,-28){\makebox(0,0)[lb]{\smash{\SetFigFont{10}{12.0}{\familydefault}{\mddefault}{\updefault}F}}}
\put(5955,-1505){\makebox(0,0)[lb]{\smash{\SetFigFont{10}{12.0}{\familydefault}{\mddefault}{\updefault}S}}}
\put(786,-643){\makebox(0,0)[lb]{\smash{\SetFigFont{10}{12.0}{\familydefault}{\mddefault}{\updefault}gaugino}}}
\put(2817,-643){\makebox(0,0)[lb]{\smash{\SetFigFont{10}{12.0}{\familydefault}{\mddefault}{\updefault}gaugino}}}
\put(4109,-643){\makebox(0,0)[lb]{\smash{\SetFigFont{10}{12.0}{\familydefault}{\mddefault}{\updefault}scalar}}}
\put(7124,-643){\makebox(0,0)[lb]{\smash{\SetFigFont{10}{12.0}{\familydefault}{\mddefault}{\updefault}scalar}}}
\put(5610,341){\makebox(0,0)[lb]{\smash{\SetFigFont{10}{12.0}{\familydefault}{\mddefault}{\updefault}fermion}}}
\put(1300,-275){\makebox(0,0)[lb]{\smash{\SetFigFont{10}{12.0}{\familydefault}{\mddefault}{\updefault}$l_{\sm{M}}(q_{\sm{M}})$}}}
\put(2400,-275){\makebox(0,0)[lb]{\smash{\SetFigFont{10}{12.0}{\familydefault}{\mddefault}{\updefault}$\bar{l}_{\sm{M}}(\bar{q}_{\sm{M}})$}}}
\put(1240,-1190){\makebox(0,0)[lb]{\smash{\SetFigFont{10}{12.0}{\familydefault}{\mddefault}{\updefault}$\psi_{l}(\psi_{q})$}}}
\put(2420,-1190){\makebox(0,0)[lb]{\smash{\SetFigFont{10}{12.0}{\familydefault}{\mddefault}{\updefault}$\bar{\psi}_{l}(\bar{\psi}_{q})$}}}
\put(4884,-643){\makebox(0,0)[lb]{\smash{\SetFigFont{10}{12.0}{\familydefault}{\mddefault}{\updefault}}}}
\put(6349,-631){\makebox(0,0)[lb]{\smash{\SetFigFont{10}{12.0}{\familydefault}{\mddefault}{\updefault}}}}
\put(5315,-324){\makebox(0,0)[lb]{\smash{\SetFigFont{10}{12.0}{\familydefault}{\mddefault}{\updefault}}}}
\put(6091,-238){\makebox(0,0)[lb]{\smash{\SetFigFont{10}{12.0}{\familydefault}{\mddefault}{\updefault}}}}
\put(5241,-1099){\makebox(0,0)[lb]{\smash{\SetFigFont{10}{12.0}{\familydefault}{\mddefault}{\updefault}}}}
\put(6300,-1160){\makebox(0,0)[lb]{\smash{\SetFigFont{10}{12.0}{\familydefault}{\mddefault}{\updefault}}}}
\end{picture}
\caption
{ Typical diagrams inducing masses for the MSSM sparticles.  Messenger
fields run in loops.  }
\label{gc}
\end{figure}

The MGMM contains, in addition to MSSM particles, two messenger
multiplets $Q_{\sm{M}}$ and $\bar{Q}_{\sm{M}}$ which belong to ${\bf
5}$ and ${\bf \bar{5}}$ representations of $SU(5)$.  We will consider
also an alternative model with the antisymmetric messengers (${\bf
10}$ and ${\bf\bar{10}}$).  Other representations  
are incompatible with asymptotic freedom
of the unified theory. Messengers couple to a MSSM singlet $\Xi$ through the
superpotential term
  
\begin{equation}
\label{scalar}
{\cal W}_{ms} = \lambda \Xi Q_{\sm{M}}\bar{Q}_{\sm{M}}.
\end{equation}
$\Xi$ obtains non-vanishing vacuum expectation values $F$ and $S$
via hidden-sector interactions,
$$\Xi = S+F\theta \theta. $$
Gauginos and scalar particles of MSSM obtain masses in one and two 
loops 
respectively (see Fig.~\ref{gc}). 
Their values are \cite{dine-f}
\begin{equation}
\label{gaugin}
M_{\lambda_{\sm{i}}} = c_i\frac{\alpha_i}{4\pi}\Lambda f_1(x)
\end{equation}
for gauginos and
\begin{equation}
\label{scalmas}
\tilde{m}^2=2\Lambda^2\sum_{i=1}^3c_iC_i\l\frac{\alpha_i}
{4\pi}\r^2f_2(x)
\end{equation}
for scalars. Here $\alpha_i$ are gauge coupling constants of
$SU(3)\times SU(2)\times U(1)$,
$C_i$ are values of the quadratic Casimir operator for various
matter fields: 
$C_3=4/3$ for color triplets (zero for singlets),
 $C_2=3/4$ for weak doublets (zero for singlets),
$C_1=\l\frac{Y}{2}\r^2$, where $Y$ is the weak
 hypercharge. For messengers belonging to the fundamental and antisymmetric
representation
one has $c_1=5/3,~c_2=c_3=1$ and $c_1=5,~c_2=c_3=3$, respectively.   

The two parameters entering eqs.~(\ref{gaugin}) and
 (\ref{scalmas}) are
$$\Lambda=\frac{F}{S}$$ and 
$$x=\frac{\lambda F}{\lambda^2S^2}$$ 
The dependence of the soft masses on $x$ is very mild, as the functions
$f_1(x)$ and $f_2(x)$ do not deviate much from 1~\cite{Dimopoulos,Martin}, 
$$
f_1(x)=\frac{1}{x^2}\left[(1+x)\ln (1+x)+(1-x)\ln (1-x)\right],
$$
$$
f_2(x)=\frac{1+x}{x^2}\left[\ln(1+x)-2 \makebox{Li}_2\!\left(\frac{x}{1+x}\right)+
\frac{1}{2}\makebox{Li}_2\!\left(\frac{2x}{1+x}\right)\right]+(x \to -x).
$$
Hence, in the absence of mixing between messengers and leptons (and quarks),
the predictions of this theory at realistic energies are determined 
predominantly by the value of $\Lambda$, while the value of $x$ is almost 
unimportant.

Unlike the masses of MSSM particles, the masses of messenger
fields strongly depend on $x$. Namely, the vacuum expectation
value of $\Xi$ mixes scalar components of messenger fields and
gives them masses
$$
M_{\pm}^2=\frac{\Lambda^2}{x^2}(1\pm x)
$$ 
It is clear that $x$ must be smaller than $1$.
Masses of fermionic components of messenger superfields
are all equal to $\frac{\Lambda}{x}$. 

In fact, the values of $x$ entering eqs.~(\ref{gaugin}) and
(\ref{scalmas}) are different for strongly and weakly interacting
sparticles. The reason is that the Yukawa couplings run differently below the 
GUT scale.
The value of $\Lambda$ remains
universal for different matter fields~\cite{Dimopoulos}. It has been found
that
the difference between 'strong' $x$ and 'weak' $x$ does not
exceed 30\%~\cite{1.3}. This effect is not essential for the values of
$x$ not very close to $1$ and  we will ignore it in what follows.

It has been argued in Ref.\cite{borzumati} that $\Lambda$ must be
larger than $70$~TeV, otherwise the theory would generically be
inconsistent with mass limits from LEP.  The characteristic features
of the model without messenger-matter mixing 
are large $\tan{\beta}$~\cite{kolda} (an estimate of
Refs.\cite{1.3,borzumati} is $\tan \beta\gtrsim 50$) and large squark masses.
Parameter $\mu$ of the Higgs sector is predicted to be about $500$ GeV. 
There is large mixing between $\tilde{\tau}_{{\sm R}}$ and
$\tilde{\tau}_{{\sm L}}$, proportional to $\tan{\beta}$ and $\mu$.
It results in the mass splitting of $\tau$-sleptons so that 
the Next Lightest Supersymmetric
Particle (NLSP) is a combination of $\tilde{\tau}_{{\sm R}}$ and
$\tilde{\tau}_{{\sm L}}$ \cite{kolda,borzumati}, the LSP being
gravitino. 
Bino is slightly
heavier, but lighter than $\tilde{\mu}_{{\sm L,\sm R}}$ and
$\tilde{e}_{{\sm L,\sm R}}$.

Messenger fields may be odd or even under R-parity. For instance,
fundamental messengers, depending on their R-parity, 
have the same quantum numbers as either
fundamental matter or fundamental Higgs,
so the messenger-matter mixing arises naturally. In the latter case
triplet messenger fields will give rise to fast proton decay due
to the possible Higgs-like mixing with ordinary fields, unless the
corresponding Yukawa couplings are smaller than
$10^{-21}$~\cite{R-even}. Another way to solve this problem is to
assume messenger doublet-triplet splitting. In that case one can try
to identify messengers with Higgs fields. Such theories were discussed
in Refs.\cite{Dvali_Shifman,Zurab}; it was shown that there are 
serious difficulties with particle spectrum and naturalness.

We will consider messengers which are odd under
R-parity. Then the components of the fundamental messengers
$Q_{\sm{M}}^{(5)}=(l_{\sm{M}},q_{\sm{M}})$ have the same quantum
numbers as left leptons and  right down-quarks, while components of
antisymmetric messengers $Q_{\sm{M}}^{(10)}$ have quantum numbers of
the right leptons, left quarks and right
 up-quarks.
We assume that there is one generation of messengers, fundamental 
or antisymmetric, and consider their mixing with
ordinary matter separately.

In the case of fundamental representation one can introduce messenger-matter
mixing~\cite{dinem}
\begin{equation}
\label{y-term}
{\cal W}_{mm}^{(5)} = H_{\sm D}L_{\hat{i}}Y^{(5)}_{\hat{i}j}E_j+H_{\sm D}
D_{\hat{i}}X^{(5)}_{\hat{i}j}Q_j
\end{equation}
where 
$$
H_{\sm D} = (h_{\sm D},\chi_{\sm D}),~~~~~~~
H_{\sm U} = (h_{\sm U},\chi_{\sm U})
$$
are Higgs doublet superfields,
\begin{displaymath}
L_{\hat{i}} = (\tilde{l}_{\hat{i}},l_{\hat{i}}) =
 \left\{ \begin{array}{ll}
(\tilde{l}_{{\sm L}\hat{i}}, l_{{\sm L}\hat{i}}) & 
\textrm{, $\hat{i}=1,..,3$}\\
(\tilde{l}_{\sm{M}}, l_{\sm{M}}) & \textrm{, $\hat{i}=4$}
\end{array} \right\},
~~~~D_{\hat{i}} = (\tilde{d}_{\hat{i}},d_{\hat{i}}) =
 \left\{ \begin{array}{ll}
(\tilde{d}_{{\sm R}\hat{i}}, d_{{\sm R}\hat{i}}) & 
\textrm{, $\hat{i}=1,..,3$}\\
(\tilde{d}_{\sm M}, d_{\sm{M}}) & \textrm{, $\hat{i}=4$}
\end{array} \right\}
\end{displaymath}
are left doublet superfields and right triplet superfields  and
\begin{displaymath}
E_{j} = (\tilde{e}_{{\sm R}j},e_{{\sm R}j}),
~~~~~~~Q_{j}=(\tilde{q}_{{\sm L}j},q_{{\sm L}j})=\l{\tilde{u}_{{\sm
L}j} \choose \tilde{d}_{{\sm
L}j}},{u_{{\sm
L}j} \choose d_{{\sm
L}j}} \r\;,~~~j=1,2,3. 
\end{displaymath}
are right lepton singlet superfields and left quark doublet
superfields, respectively.

Hereafter $\hat{i},\hat{j}=1,..,4$ label the three left
lepton (and right down-quark) generations and the messenger field, $i,j=1,..,3$
correspond to the three leptons (quarks) and 
$Y^{(5)}_{\hat{i}j},X^{(5)}_{\hat{i}j}$ are the
$4\times3$ matrices of Yukawa couplings,
$$
Y_{\hat{i}j}^{(5)} = 
\left(
\begin{array}{ccc}
Y_{e}& 0 & 0\\
0 & Y_{\mu} & 0\\
0 & 0 & Y_{\tau}\\
Y_{41}^{(5)}& Y_{42}^{(5)}& Y_{43}^{(5)}
\end{array}
\right),~~~~
X_{\hat{i}j}^{(5)} = 
\left(
\begin{array}{ccc}
Y_d& 0 & 0\\
0 & Y_s & 0\\
0 & 0 & Y_b\\
X_{41}^{(5)}& X_{42}^{(5)}& X_{43}^{(5)}
\end{array}
\right)
$$
In terms of component fields, the tree level scalar potential 
and
Yukawa terms are
\begin{eqnarray}
\label{scalss}
V^{(5)} & = & \lambda^2S^2 \tilde{l}_{\sm M}^*\tilde{l}_{\sm M} +
\mu^2h_{{\sm D}}h_{{\sm D}}^* +
 |\lambda S\tilde{\bar{l}}_{\sm M} + h_{{\sm D}}\tilde{e}_{{\sm
R}j}Y^{(5)}_{4j}|^2 + |\mu h_{{\sm U}} +
Y^{(5)}_{\hat{i}j}\tilde{l}_{\hat{i}}\tilde{e}_{{\sm
R}j}+X_{\hat{i}j}^{(5)}\tilde{d}_{\hat{i}}\tilde{q}_{{\sm L}j}|^2
\nonumber \\
& & + |Y^{(5)}_{ij}h_{{\sm D}}\tilde{e}_{{\sm R}j}|^2 + 
|Y^{(5)}_{\hat{i}j}\tilde{l}_{\hat{i}}h_{{\sm D}}|^2 +
 \Bigl( \lambda Sl_{\sm M}\bar{l}_{\sm M}
 - \lambda F\tilde{l}_{\sm M}\tilde{\bar{l}}_{\sm M}   \nonumber \\& & + 
Y^{(5)}_{\hat{i}j}\bigl( h_{{\sm D}}
l_{\hat{i}}e_{{\sm R}j} + 
\chi_{{\sm D}}\tilde{l}_{\hat{i}}
e_{{\sm R}j} + 
\chi_{{\sm D}}l_{\hat{i}}\tilde{e}_{{\sm R}j}\bigr)
 + \mu\chi_{{\sm D}}\chi_{{\sm U}} + h.c. \Bigr) \\& &
 +\lambda^2S^2\tilde{q}_{\sm M}^*\tilde{q}_{\sm M} + |\lambda
S\tilde{\bar{q}}_{\sm M} + 
h_{{\sm D}}\tilde{q}_{{\sm L}j}X^{(5)}_{4j}|^2 +  
 |X^{(5)}_{ij}h_{{\sm D}}\tilde{q}_{{\sm L}j}|^2 + 
|X_{\hat{i}j}^{(5)}\tilde{d}_{\hat{i}}h_{{\sm D}}|^2 \nonumber\\& & +
 \Bigl( \lambda Sq_{\sm M}\bar{q}_{\sm M}
 - \lambda F \tilde{q}_{\sm M}\tilde{\bar{q}}_{\sm M}+  
X^{(5)}_{\hat{i}j}\bigl( h_{{\sm D}}
d_{\hat{i}}q_{{\sm L}j} + 
\chi_{{\sm D}}\tilde{d}_{\hat{i}}
q_{{\sm L}j} + 
\chi_{{\sm D}}d_{\hat{i}}\tilde{q}_{{\sm L}j}\bigr)
 + h.c. \Bigr),\nonumber
\end{eqnarray}
where $\mu$ is the usual parameter of MSSM.
Besides these terms, there are soft-breaking 
terms coming
from loops involving messenger fields.
In the absence of messenger-matter mixing~(\ref{y-term})
 they have the form (at the SUSY breaking scale, which is of
 order of $\Lambda$) 
\begin{eqnarray}
\label{scalsf} 
{\cal V}_{sb} & = & \tilde{m}_{{\sm L}i}^2\tilde{e}_
{{\sm L}i}\tilde{e}_{{\sm L}i}^* 
+ \tilde{m}_{{\sm R}i}^2 \tilde{e}_{{\sm R}i}\tilde{e}_
{{\sm R}i}^* +\nonumber \\
 & & \tilde{m}_{{\sm{\sm q}}{\sm L}i}^2\tilde{q}_
{{\sm L}i}\tilde{q}_{{\sm L}i}^* 
+ \tilde{m}_{{\sm{\sm q}}{\sm R}i}^2 \tilde{q}_{{\sm R}i}\tilde{q}_
{{\sm R}i}^* + 
\tilde{m}_{{\sm{\sm d}}{\sm L}i}^2\tilde{d}_
{{\sm L}i}\tilde{d}_{{\sm L}i}^* 
+ \tilde{m}_{{\sm{\sm d}}{\sm R}i}^2 \tilde{d}_{{\sm R}i}\tilde{d}_
{{\sm R}i}^*.
\end{eqnarray}
where $\tilde{m}_{{\sm L}j}^2$, $\tilde{m}_{{\sm R}j}^2$, 
$\tilde{m}_{u{\sm L}j}^2$, 
$\tilde{m}_{ d{\sm L}j}^2$, $\tilde{m}_{u{\sm R}j}^2$, 
$\tilde{m}_{d{\sm R}j}^2$ are
given by eq.~(\ref{scalmas}). Low energy effective potential can be
obtained from eq.~(\ref{scalsf}) by making use of renormalization
group equations. It is worth noting that the boundary conditions for
soft Higgs mass term $B\mu h_{\sm U}h_{\sm D}$ and for scalar
trilinear couplings are set to zero in MGMM because their values at
the scale $\Lambda$ are suppressed in comparison with other terms in 
eq.~(\ref{scalsf}). Messenger-matter mixing modifies
 eq.~(\ref{scalsf}); we will consider this modification later on.

There is no CP-violation in this
theory in lepton sector. 
Arbitrary phases may be rotated away by 
redefinition of the lepton fields. On the other hand, there appears 
CP-violation in quark sector, in  addition to the CKM mechanism. There
are three phases in the matrix $X^{(5)}$ and only one of them 
may be set equal to zero by redefinition of the messenger
fields. 

The same formalism applies to antisymmetric
messenger fields. The mixing terms have the following form, 

\begin{equation}
\label{y-term-10}
{\cal W}_{mm}^{(10)} = H_{\sm D}E_{\hat{i}}Y^{(10)}_{\hat{i}j}L_j+H_{\sm D}
Q_{\hat{i}}W^{(10)}_{\hat{i}j}D_j +
H_{\sm U}
U_{\hat{i}}X^{(10)}_{\hat{i}\hat{j}}Q_{\hat{j}}
\end{equation}
One can make the first term in the Lagrangian (\ref{y-term-10}) real
by redefinition of lepton fields. So, CP-violation
comes only from the last two terms. In addition to the SM CP-violating
phase they contain six new phases in 
Yukawa couplings\footnote{The seventh phase in
$X^{(10)}_{44}$ which corresponds to messenger-messenger
mixing will not be of interest because its effects on
the processes in SM sector are negligible.} $X_{4j}^{(10)}$ and
$X_{j4}^{(10)}$ and two phases in couplings 
$W_{4j}^{(10)}$.

To summarize, messenger-matter mixing in the leptonic sector occurs
through the Yukawa couplings $Y_{4i}^{(5)}$ or  $Y_{4i}^{(10)}$
($i=1,2,3$), depending on the representation of the messenger fields.
Likewise, the mixing in the quark sector appears through 
$X_{4i}^{(5)}$ or
 $X_{i4}^{(10)}$,
 $X_{4i}^{(10)}$,
 $W_{4i}^{(10)}$. In the following sections we
sometimes use the collective notation $Y_i$ for the 
 couplings 
$Y_{4i}$, $X_{4i}$, $X_{i4}$ and $W_{4i}$ in 
statements applicable to all of them.

\section{Induced mixing of matter fields.}
In this section we consider mixing between the fields of MSSM that appears 
after messengers are integrated out.
It is straightforward to check that fermion mixing terms are small at the tree level 
(the tree level fermion mass matrices are presented in Appendix {\bf
A}). In principle, this mixing (the off-diagonal
terms in eqs.~(\ref{ufl}) and (\ref{ufr})) may lead to lepton and quark
flavor violation due to one loop diagrams involving scalars and
gauginos~\cite{barbieri}.  However, these mixing terms are
negligible due to see-saw type mechanism: in MGMM one definitely
has $\lambda S>10^4$ GeV, $\tan \beta \gtrsim 1$ and 
the tree level fermion mixing terms are smaller than $10^{-4}$
 even at $Y_i\sim 1$,
see eqs.~(\ref{ufl}) and (\ref{ufr}). The corresponding
contributions to flavor violating rates are too small to
be observable. 

Mixing in the scalar sector is also small at the tree level (the tree
level mass matrices of scalars are given in Appendix {\bf B}).  
After the scalar messengers are integrated out at the tree level,
the lepton flavor violating terms in the mass matrix of right
sleptons and quark flavor violating terms in the mass matrix of 
down-squarks are of order 
\begin{equation}
\label{drev}
Y_i^*Y_j\l v_{{\sm D}}^2x^2+\frac{\mu ^2
v_{{\sm U}}^2x^2}{\Lambda ^2}\r
\end{equation}
for generic values of $x$ (not too 
close to 1). Analogous expression with interchange of $v_{\sm{U}}$
and
$v_{\sm{D}}$ holds for up-squarks in the case of antisymmetric 
messengers.  These terms are smaller than
the one loop contributions (see below). The only
substantial non-diagonal terms in the slepton mass matrix are $(- \lambda
F)$ in the messenger sector and $\tilde{\tau}_{{\sm
R}}-\tilde{\tau}_{{\sm L}}$ mixing proportional to $\tan{\beta}$. 
Due to the latter, the NLSP
is $\tilde{\tau}$ at $\tan{\beta}>25$~\cite{kolda}, 
while at lower $\tan{\beta}$ the NLSP is neutralino. The only
substantial non-diagonal terms in the squark mass matrix are $(- \lambda
F)$ in the messenger sector and $\tilde{b}_{{\sm
R}}-\tilde{b}_{{\sm L}}$  mixing (and $\tilde{t}_{{\sm
R}}-\tilde{t}_{{\sm L}}$ mixing in the case of antisymmetric messengers).  

The dominant contributions to mixing in slepton and squark sectors 
appear through one loop diagrams from trilinear terms in the
superpotential, that involve $H_{\sm D}$ for fundamental messengers,
and also $H_{\sm U}$ for antisymmetric ones.  The fact that the one
loop mixing terms of scalars are proportional to the large
parameter $\Lambda^2$ is obvious from eq.~(\ref{scalss}): say, one
of the cubic terms in the scalar potential, $[\lambda S
Y^{(5)}_{4j}\bar{l}_{\sm M}^*h_{\sm D}\tilde{e}_{{\sm R}j}+$ h.c.], contains
$(\lambda S)=\frac{\Lambda}{x}$ explicitly.

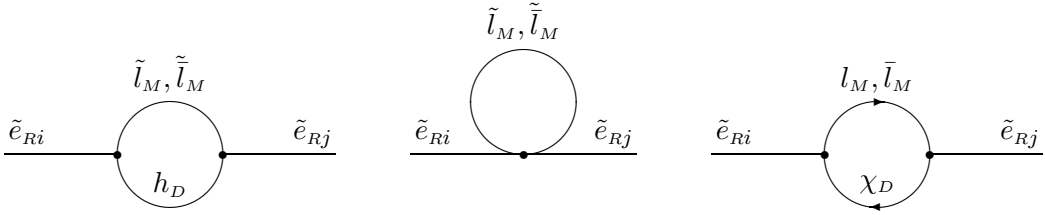
\begin{figure}[htb]
\begin{center}
\unitlength=1.00mm
\special{em:linewidth 0.4pt}
\linethickness{0.4pt}
\begin{picture}(148.00,40.00)(12,100)
\put(10.00,120.00){\line(1,0){15.00}}
\put(32.00,120.00){\circle{14.00}}
\put(39.00,120.00){\line(1,0){15.00}}
\put(64.00,120.00){\line(1,0){30.00}}
\put(79.00,127.00){\circle{14.00}}
\put(13.00,122.00){\makebox(0,0)[cb]{$\tilde{e}_{{\sm R}i}$}}
\put(51.00,122.00){\makebox(0,0)[cb]{$\tilde{e}_{{\sm R}j}$}}
\put(67.00,122.00){\makebox(0,0)[cb]{$\tilde{e}_{{\sm R}i}$}}
\put(91.00,122.00){\makebox(0,0)[cb]{$\tilde{e}_{{\sm R}j}$}}
\put(32.00,129.00){\makebox(0,0)[cb]{$\tilde{l}_{\sm M},\tilde{\bar{l}}_{\sm M}$}}
\put(32.00,116.00){\makebox(0,0)[cc]{$h_{\sm D}$}}
\put(79.00,136.00){\makebox(0,0)[cb]{$\tilde{l}_{\sm M},\tilde{\bar{l}}_{\sm M}$}}
\put(104.00,120.00){\line(1,0){15.00}}
\put(126.00,120.00){\circle{14.00}}
\put(133.00,120.00){\line(1,0){15.00}}
\put(107.00,122.00){\makebox(0,0)[cb]{$\tilde{e}_{{\sm R}i}$}}
\put(145.00,122.00){\makebox(0,0)[cb]{$\tilde{e}_{{\sm R}j}$}}
\put(126.00,129.00){\makebox(0,0)[cb]{$l_{\sm M},\bar{l}_{\sm M}$}}
\put(126.00,116.00){\makebox(0,0)[cc]{$\chi_{\sm D}$}}
\put(126.00,127.00){\vector(1,0){1.00}}
\put(126.00,113.00){\vector(-1,0){1.00}}
\put(79.00,120.00){\circle*{1.00}}
\put(39.00,120.00){\circle*{1.00}}
\put(25.00,120.00){\circle*{1.00}}
\put(119.00,120.00){\circle*{1.00}}
\put(133.00,120.00){\circle*{1.00}}
\end{picture} 
\caption{The diagrams dominating the slepton
mixing matrix.}
\label{gopa}
\end{center}
\end{figure}

After diagonalizing the messenger mass matrix we obtain the
diagrams contributing to slepton mixing to the order $(\lambda
S)^2$, which are shown in Fig.~\ref{gopa}.
Similar diagrams contribute to the mass matrix
for left squarks. In the antisymmetric case we obtain from
similar diagrams mixing terms in the mass matrices of left sleptons and
squarks. 
If supersymmetry were unbroken, the sum of these diagrams would be equal to
zero.
In our case of broken supersymmetry the resulting
contributions to the mass matrices of MSSM scalars are
 \newline \newline 
\begin{eqnarray}
\label{delta-mixing-1}
{\bf for~~fundamental~~messengers}\nonumber\\
right~~sleptons~~~~\delta m^2_{ij}=-
\frac{\Lambda^2}{8\pi^2}Y^{(5)}_{4i}Y^{(5)}_{4j}f_3(x)\\
\label{delta-mixing-2}
left~~(up~~and~~down)~~squarks~~~~\delta m^2_{ij}=-
\frac{\Lambda^2}{16\pi^2}X^{(5)*}_{4i}X^{(5)}_{4j}f_3(x)\\
{\bf for~~antisymmetric~~messengers}\nonumber\\
\label{delta-mixing-3}
left~~sleptons~~(selectrons~~and~~sneutrino)~~~\delta m^2_{ij}=-
\frac{\Lambda^2}{16\pi^2}Y^{(10)}_{4i}Y^{(10)}_{4j}f_3(x)\\
\label{cpviol_1}
right~~up~~squarks~~~\delta m^2_{ij}=-
\frac{\Lambda^2}{8\pi^2}X^{(10)*}_{i4}X^{(10)}_{j4}f_3(x)\\
\label{cpviol_2}
left~~(up~~and~~down)~~squarks~~~\delta m^2_{ij}=-
\frac{\Lambda^2}{16\pi^2}X^{(10)*}_{4i}X^{(10)}_{4j}f_3(x)\\
\label{delta-mixing-4}
right~~down~~squarks~~~~\delta m^2_{ij}=-
\frac{\Lambda^2}{8\pi^2}W^{(10)*}_{4i}W^{(10)}_{4j}f_3(x)
\end{eqnarray}
where 
\begin{equation}
\label{f3}
f_3(x)=\frac{1}{x^2}\Big\{ 
-\ln{(1-x^2)}-\frac{x}{2}
 \ln{\Big(\frac{1+x}{1-x}\Big)}\Big\}
\end{equation}
These terms were obtained to the zeroth order in the Higgs masses. The higher 
order
contributions are suppressed by the squared gauge coupling
constants or by small ratio $\frac{\mu}{\Lambda}$. Notice that mixing
terms~(\ref{delta-mixing-2}), (\ref{cpviol_1})-(\ref{delta-mixing-4}) 
violate CP. 

Since scalars get negative shifts in squared masses, these expressions for the
soft terms immediately imply theoretical bounds on Yukawa
couplings $Y_i$ which come from the requirement \cite{dinem}
that none of the scalar squared masses becomes negative (see below).

Now, let us see that the  
contributions~(\ref{delta-mixing-1})~--~(\ref{delta-mixing-4}) are
much larger than the tree level mixing terms. As an example,
at small $x$ one has for the one loop terms
\begin{equation}
\delta m^2_{ij}\simeq - \frac{a}{6}\frac{\Lambda^2}{16\pi^2}
Y^*_iY_jx^2 
\label{small_x}
\end{equation}
where $a =1~~\mbox{or}~~2$ depending on the $SU(2)$ representation of
the fields. These terms dominate over
the tree level ones (\ref{drev}) provided the following inequalities
are satisfied,  
$$
v_{\sm D}^2+\frac{\mu^2v_{\sm
U}^2}{\Lambda^2}<\frac{\Lambda^2}{96\pi^2}
\;,~~~
v_{\sm U}^2+\frac{\mu^2v_{\sm D}^2}{\Lambda
^2}<\frac{\Lambda^2}{96\pi^2}\;.
$$
In MGMM one has $\Lambda>10$~TeV and $\mu\simeq 400\div
500$~GeV, so these inequalities indeed hold.

In the case of very small values of $x$
the dominant contributions to $\delta m_{ij}^2$ come from two loops
rather than one loop. The two-loop
contributions appear from the
same diagrams as those shown in Fig.~\ref{gopa} but with one-loop enhanced
vertices. The result for the diagonal mass terms was found in Ref.~\cite{GR} 
and has the following form, 
\begin{equation}
\delta m_i^2=\frac{d_i}{8\pi^2}\frac{|Y_{4i}|^2}{4\pi}\l 
\frac{D}{2}\frac{|Y_{4i}|^2}{4\pi}-C\alpha\r\Lambda^2.
\label{delta_2}
\end{equation}
Here $C=\sum C_i$ and $D=\sum d_i$ where the sums run over all
superfields participating in the mixing interactions; $C_i$
denote quadratic Casimir operators and $d_i$ are the numbers of fields
circulating in the Yukawa loop. The off-diagonal terms are similar. 
One can compare this result 
to eqs.~(\ref{delta-mixing-1})~--~(\ref{delta-mixing-4}) and find the
region of parameters where the contributions~(\ref{delta_2}) become
significant, 
\begin{equation}
x^2<\frac{3}{\pi}\frac{d_i}{a}\left|\frac{D}{2}\frac{|Y_{4i}|^2}
{4\pi}-C\alpha\right|
\label{region}
\end{equation} 
Requirement of positivity of the scalar masses implies that the first
term in eq.~(\ref{region}) has to be smaller than the second
one, so the two-loop corrections become essential in the region
$x\sim\sqrt{\alpha}$.

\section{Electroweak breaking and squark masses}
It was already mentioned  that radiative electroweak breaking in
MGMM without messenger-matter mixing leads to  large values of $\tan{\beta}\gtrsim 50$.
  The usual way to avoid this limit is to
assume some extra soft contribution to the Higgs sector of the theory.
In this section we consider electroweak breaking in the model with 
messenger-matter mixing. In particular, we show that wide range 
of values of 
$\tan{\beta}$ is now allowed without any additional parameters in the Higgs sector
of the model. For definiteness, we concentrate  on the
case of fundamental messengers. 

Minimization of the Higgs potential results in the following two
equations, 
\begin{eqnarray}
\label{ewb}
\sin{2\beta}=\frac{-2B\mu}{m^2_{h_{\sm U}}+m^2_{h_{\sm D}}+2\mu^2}\nonumber\\
\mu^2=\frac{m^2_{h_{\sm D}}-m^2_{h_{\sm U}}\tan^2{\beta}}{\tan^2{\beta}-1}-\frac{1}{2}M_Z^2
\end{eqnarray}
The parameter $B$ characterizes the magnitude of the soft 
mixing term in the Higgs sector, $B\mu h_{{\sm U}}h_{{\sm D}}$. 
At the two loop level it is equal to~\cite{kolda}
\begin{equation}
\label{B-term}
B=M_{\lambda_2}(-0.12+0.17Y_t^2), 
\end{equation}
where $M_{\lambda_2}$ is given by eq.~(\ref{gaugin}). 

In MGMM without messenger-matter mixing the value of the soft mass
$m^2_{h_{\sm D}}$ is given by eq.~(\ref{scalmas}) while $m^2_{h_{\sm
U}}$ receives additional large negative one-loop correction due to
large Yukawa coupling between $H_{\sm U}$ and t-quark,
\begin{equation}
\label{toploop}
\delta m^2_{h_{\sm U}}=-\frac{3Y^2_t}{4\pi^2}m^2_{\tilde{t}}\ln\left(\frac
{\Lambda}{xm_{\tilde{t}}}\right)\;.
\end{equation}
It is natural to expect large value of  $\tan{\beta}$ in such 
a situation. Indeed, if the two Higgs fields were not mixed
and only one of them ($h_{\sm U}$ in our case) obtained the negative mass
squared  then $\tan{\beta}\equiv \frac{v_{\sm U}}{v_{\sm D}}$ would
be equal to infinity. Of course, there is mixing between $h_{\sm U}$ and $h_{\sm D}$
in SUSY theories due to $\mu$-term. However, it follows from
eqs.~(\ref{scalmas}) and (\ref{toploop}) that 
$\delta m^2_{h_{\sm U}}\gg m^2_{h_{\sm U}},m^2_{h_{\sm D}}$ so
that eq.~(\ref{ewb}) takes the following simple form
\begin{equation}
\mu^2\simeq -\delta m^2_{h_{\sm U}},~~~~\sin{2\beta}\simeq -
\frac{2B}{\mu}
\label{large}
\end{equation}
It is clear from eqs.~(\ref{B-term}) and (\ref{large}) that $\sin 2\beta\ll 1$,
and, therefore, $\tan\beta\gg 1$. This simple estimate gives
\begin{equation}
\tan{\beta}\gtrsim 50
\label{50}
\end{equation}
At so large values of $\tan{\beta}$ other
corrections to Higgs masses (e.g., the corrections to $m^2_{h_{\sm
D}}$ due to the interaction with
b-quark analogous to eq.~(\ref{toploop})) have to be taken into account. However, 
detailed analysis~\cite{1.3,borzumati} in which these corrections are
included, confirms the estimate~(\ref{50}). 

In the presence of messenger-matter mixing, $m_{h_{\sm D}}^2$ receives
additional negative contributions from the diagrams analogous to those
shown in Fig.~\ref{gopa}. The purpose of this section is to
demonstrate that these contributions can make $\tan{\beta}$ to be as
low as $\tan{\beta}\sim 1$. The simplified analysis along the lines
outlined above is sufficient for this purpose, as additional
contributions (e.g., due to b-quark) are negligible at not too large 
$\tan{\beta}$. Certainly, our
results become qualitative at $\tan{\beta}\gtrsim 30$, when these
corrections are significant. 

\begin{figure}[htb]
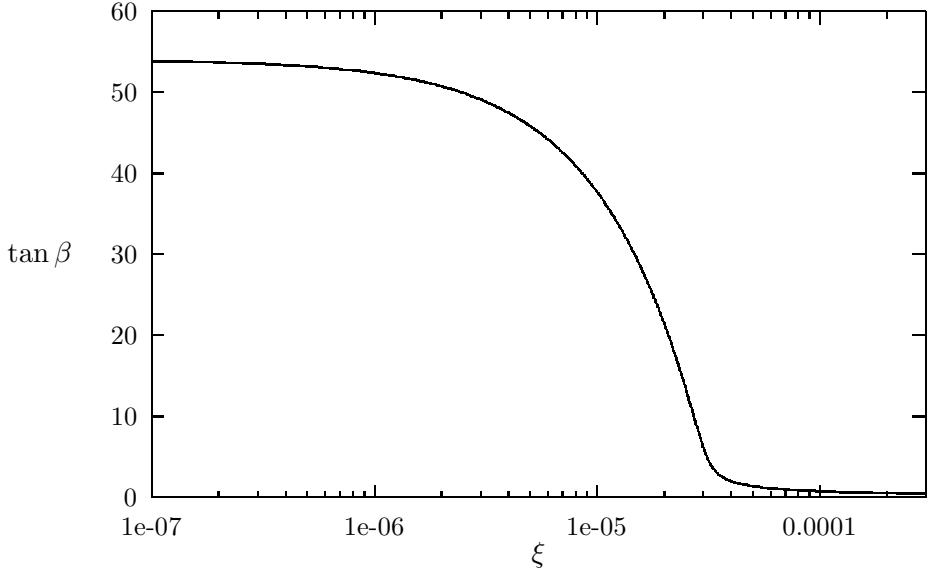

% GNUPLOT: LaTeX picture
\setlength{\unitlength}{0.240900pt}
\ifx\plotpoint\undefined\newsavebox{\plotpoint}\fi

\caption
{$\tan\beta$ as function of $\xi=\delta 
m^2_{h_{\sm D}}/\Lambda^2=-d_{\sm D}^{(5)}\frac{f_3(x)}{16\pi^2}$.}
\label{tangent-mixing}
\end{figure}

In our case of fundamental messengers, the contribution to $m_{h_{\sm
D}}^2$ due to messenger-matter mixing is equal to
\begin{equation}
\label{dloop}
\delta m^2_{h_{\sm D}}=-d^{(5)}_{\sm
D}\frac{\Lambda^2}{16\pi^2}f_3(x)\;,
\end{equation}
where
\begin{equation}
\label{parameter}
d^{(5)}_{\sm D}=\sum_{i=1}^3\left(|Y_{4i}^{(5)}|^2+3|X_{4i}^{(5)}|^2\right) 
\end{equation}
Depending on the Yukawa coupling constants, the mass splitting $\delta
m^2_{h_{\sm D}}$ may be of the same order as $\delta m^2_{h_{\sm
U}}$. Therefore the value of $\sin{2\beta}$ gets modified as compared
to eq.~(\ref{large}). Instead, one has the estimate,
\begin{equation}
\mu^2\simeq\delta m^2_{h_{\sm U}},~~~~\sin 2\beta\simeq 
\frac{-2B\mu}{\mu^2-\delta m^2_{h_{\sm D}}}
\label{small}
\end{equation}
Therefore, messenger-matter mixing reduces $\tan\beta$. 
The result of numerical solution of eq.~(\ref{ewb}) in
the theory with mixing is shown in Fig.~\ref{tangent-mixing}, where 
we set $x=1$ in the argument of the logarithm
in eq.~(\ref{toploop})\footnote{The extension to small $x$ 
is straightforward. The results for $\tan{\beta}$ change only slightly
because of weak logarithmic dependence on $x$.}. 

\begin{figure}
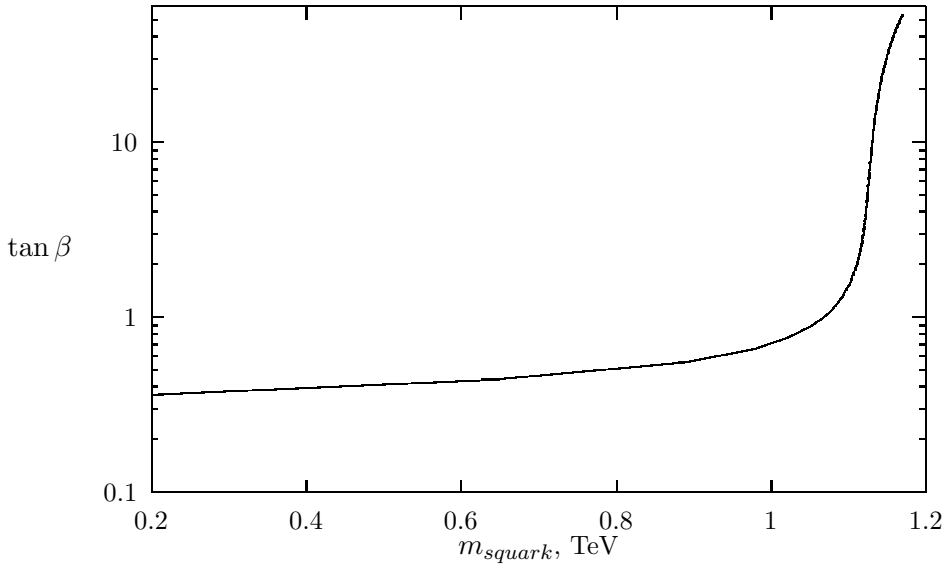

% GNUPLOT: LaTeX picture
\setlength{\unitlength}{0.240900pt}
\ifx\plotpoint\undefined\newsavebox{\plotpoint}\fi
%
\caption{$\tan\beta$ as function of the lightest squark mass $m_{squark}$.}
\label{tangent-mass}
\end{figure}

We will see in section~\ref{conyuk} that experimental bounds coming
from flavor physics constrain not the individual couplings $Y_i$ but
the products $Y_iY_j$.  So, the only bound on $\delta m^2_{h_{\sm D}}$
is theoretical one, related to the requirement of positivity of masses
of squarks and sleptons.  Namely, with loop
corrections~(\ref{delta-mixing-1})~--~(\ref{delta-mixing-4}) to the
scalar mass matrix included, two of its eigenvalues remain the same and
the third one decreases. Its value is still positive only if
\begin{equation}
\tilde{m}^2+\sum_{j=1}^{3}\delta m^2_{jj}>0\;.
\label{delta}
\end{equation}
In the case of fundamental messengers one obtains for slepton couplings
\begin{equation}
\label{Yi2}
\sum |Y^{(5)}_{4i}|^2<{80\pi^2\over 3}\frac{f_2(x)}{f_3(x)}
\l\frac{\alpha_1}{4\pi}\r^2
\end{equation}
which at small $x$ reduces to
\begin{equation}
\label{toosmall}
\sum |Y^{(5)}_{4i}|^2x^2<10^{-3}\;.
\end{equation}
In squark sector one finds
\begin{equation}
\label{Xi2}
\sum |X^{(5)}_{4i}|^2<{128\pi^2\over 3}\frac{f_2(x)}{f_3(x)}
\l\frac{\alpha_3}{4\pi}\r^2
\end{equation}
which in the case of small $x$ reduces to
$$
\sum |X^{(5)}_{4i}|^2x^2<0.1\;.
$$
These inequalities give $\delta m^2_{h_{\sm D}}/\Lambda^2<3\times
10^{-4}$. Hence, all values of $\xi$ shown in
Fig.~\ref{tangent-mixing} are allowed; the
value of $\tan{\beta}$ strongly depends on the mixing parameter $d$
and may actually be rather small. One can show that this conclusion
survives if higher order corrections are taken into account. 

On the other hand,
the parameter $\mu$ weakly depends
on mixing and grows rapidly only at $\tan{\beta}\lesssim 1$.

It is worth pointing out the correspondence between the mass spectrum
and the value of $\tan{\beta}$ in MGMM with mixing. The theoretical limit
on $Y_{4i}^{(5)}$, eq.~(\ref{toosmall}), 
implies that these couplings are always too small to
alter the value of $\tan{\beta}$ and one can neglect their
contribution in
eq.~(\ref{parameter}). Consequently,  small $\tan{\beta}$ in this 
model is correlated with large corrections to the mass matrix of left
squarks.
As was already mentioned, these corrections lead to the
significant decrease of the mass of one of the squark doublets. 

The relation between $\tan{\beta}$ and the lightest squark mass at
$\Lambda=100$~TeV is presented in Fig.~\ref{tangent-mass}.  There are
two distinct regions on this plot. The first region corresponds to 
rapidly changing $\tan\beta$ ($50\div 1$) and slowly changing
$m_{squark}$ ($1.2\div 1.0$~TeV). In fact, there exists also squark
mass splitting due to the mixing between left and right squarks
proportional to $\tan{\beta}$, which was not taken into account
above. It emerges even without messenger-matter mixing, in the same
way as $\tilde{\tau}_{\sm R}-\tilde{\tau}_{\sm L}$ mass splitting
mentioned in section~\ref{s2}, and is of order $0.1\div 0.2$~TeV at
large $\tan \beta$.  Hence, the mass range of the lightest squark
$m_{squark} \sim 1.2\div 1.0$~TeV is not a peculiarity of the model
with messenger-matter mixing.  On the other hand, the second region
($\tan{\beta}\sim 1$, $>m_{squark}=250\div 1000$~GeV, where the lower
bound is the experimental limit on the squark mass) provides a
distinctive signature of the model with messenger-matter mixing.  The
observation of one and only one pair of relatively light left up and
down squarks would be a strong indication of the gauge mediation
scenario with fundamental messengers; in this case $\tan \beta$ is
predicted to be quite low and its value is correlated to the mass of
the lightest squark.
 
In the case of antisymmetric messengers, messenger-matter mixing
contributes to $m^2_{h_{\sm D}}$ and $m^2_{h_{\sm U}}$,
\begin{equation}
\label{dloop1}
\delta m^2_{h_{\sm D}}=-d^{(10)}_{\sm
D}\frac{\Lambda^2}{16\pi^2}f_3(x)\;\;\;,\delta m^2_{h_{\sm U}}=-d^{(10)}_{\sm
U}\frac{\Lambda^2}{16\pi^2}f_3(x)\;
\end{equation}
where
$$
d^{(10)}_{\sm D}=\sum_{i=1}^3\l|Y_{4i}^{(10)}|^2+3|W_{4i}^{(10)}|^2\r\;\;\;, 
d^{(10)}_{\sm U}=\sum_{i=1}^3\l3|X_{i4}^{(10)}|^2+3|X_{4i}^{(5)}|^2\r\;.
$$
The requirement of positivity for squared scalar masses gives
\begin{eqnarray}
\sum |Y^{(10)}_{4i}|^2<72\pi^2\frac{f_2(x)}{f_3(x)}
\l\frac{\alpha_2}{4\pi}\r^2\;,\nonumber\\
\sum |W^{(10)}_{4i}|^2<64\pi^2\frac{f_2(x)}{f_3(x)}
\l\frac{\alpha_3}{4\pi}\r^2\;,\nonumber\\
\sum |X^{(10)}_{i4}|^2<64\pi^2\frac{f_2(x)}{f_3(x)}
\l\frac{\alpha_3}{4\pi}\r^2\;,\nonumber\\
\sum |X^{(10)}_{4i}|^2<128\pi^2\frac{f_2(x)}{f_3(x)}
\l\frac{\alpha_3}{4\pi}\r^2\;,\nonumber
\end{eqnarray}
which in the case of small $x$ reduce to
\begin{eqnarray}
\sum |Y^{(10)}_{4i}|^2x^2<3\cdot10^{-2}\;,\nonumber\\
\sum |W^{(10)}_{4i}|^2x^2<0.2\;,\nonumber\\
\sum |X^{(10)}_{4i}|^2x^2<0.2\;,\nonumber\\
\sum |X^{(10)}_{i4}|^2x^2<0.4\;.\nonumber\\
\end{eqnarray}
These inequalities give 
$\xi_{\sm D}=\delta m^2_{h_{\sm D}}/\Lambda^2<6\cdot10^{-4}$ and 
$\xi_{\sm U}=\delta m^2_{h_{\sm U}}/\Lambda^2<2\cdot10^{-3}$. 
The small values of $\tan\beta$ are also possible, if $\xi_{\sm D}$
is large enough so
that $m^2_{h_{\sm D}}$ and $m^2_{h_{\sm U}}$ are of the same order.
This happens when some of the Yukawa couplings $W^{(10)}_{4i}$ are
sufficiently large. Large $W^{(10)}_{4i}$ induce large corrections 
to the mass matrix of right
down squarks, $\delta\tilde{m}^2_{ij}\propto
W^{(10)}_{4i}W^{(10)}_{4j}x^2$.
Consequently, small $\tan\beta$ is correlated with light right 
down squark in the case of antisymmetric messengers. 

\section{Flavor violation}
\label{conyuk}
Let us now consider the effects of scalar mixing on the usual leptons
and quarks.  In what follows we neglect flavor violating amplitudes
coming not from slepton and squarks matrices given by
eqs.~(\ref{delta-mixing-1})~--~(\ref{delta-mixing-4}) but from
renormalization of gauge couplings by messengers. These effects
become significant at the same range of $x$ as the two-loop
corrections, eq.~(\ref{region}), and play a similar role. Namely,
they prevent $Y_i$ to be arbitrarily large at small $x$.  The point
is that all limits obtained from eqs.~(\ref{delta-mixing-1})~--~
(\ref{delta-mixing-4}) contain the
products $Y^*_iY_jf_3(x)\sim Y^*_iY_jx^2$ while two-loop corrections
(\ref{region}), as well as the corrections originating from the
gauge coupling renormalization provide limits on Yukawa couplings
themselves.

Also, we make further simplification. Various mixing terms
(including Standard Model ones) provide additive
contributions to the amplitudes of the processes under consideration.
Nevertheless, we obtain the constraints by considering every contribution
separately, i.e., by setting all others to zero and neglecting possible
interference. This will be sufficient for 
understanding the allowed magnitudes of the Yukawa couplings that induce
mixing\footnote{It worth mentioning, however, that in case of the
lepton mixing there are no Standard Model contributions and our
limits are exact in this sense.}.  We again point out that we consider
fundamental and antisymmetric cases separately, so we do not discuss
contributions which are proportional to $Y^{(5)}Y^{(10)}$ in spite of
their presence in theories with various messengers belonging to
different representations.  More accurate consideration of these
points is straightforward. 

As shown in the previous section, the wide range of $\tan{\beta}$ is
allowed in this model, depending on the mixing terms.  We will consider the
cases of high $\tan{\beta}\sim 50$ and low $\tan{\beta}\sim 1$
separately. The reason is that two different contributions
to the amplitudes dominate in these two regimes. The first
one\footnote{This contribution was missed in Ref.~\cite{we}.} is
proportional to $\tan{\beta}$ and dominates at high $\tan{\beta}$. The
second one is independent of $\tan{\beta}$ and becomes significant at
low $\tan{\beta}$.

\subsection{Lepton sector}

Let us first consider the case of fundamental messengers.  
There are two types of non-diagonal elements in the slepton
mass matrix.  The first one is the flavor diagonal
left-right mixing, coming from the tree level potential (\ref{scalss})
and proportional to $\mu m_{f}\tan{\beta}$, where $m_{f}$ is the mass of
the corresponding fermion flavor. The second type is the flavor
violating mixing~(\ref{delta-mixing-1}) in the sector of right sleptons.
  
The mass matrix of right sleptons can be diagonalized by an orthogonal
rotation; let us denote the corresponding orthogonal 
$3\times 3$ matrix by $V_{ij}$. As a
result, one of the masses of right sleptons 
(without loss of generality we denote the corresponding 
slepton as $\tilde{\tau}_{\sm{R}}$) receives negative 
contribution equal
to 
$$
\Delta m^2_{{\sm R},3}=-{\Lambda^2\over 8\pi^2}f_3(x)\Delta^2\;,
$$
where
$$
\Delta^2=(Y^{(5)}_{41})^2+(Y^{(5)}_{42})^2+(Y^{(5)}_{43})^2\;.
$$
There are two sources of flavor violation after this rotation, namely, 
the interactions between neutralino, right sleptons and leptons and left-right
mixing.
The situation here is completely  analogous to the
lepton flavor violation in the SUSY $SU(5)$ model with universal soft terms 
at the Planck scale, which was studied in detail in Ref.~\cite{barbieri}. 

The resulting rate of $\mu\to e\gamma$ decay is equal to~\cite{barbieri}
\begin{equation}
\label{muegrate}
\Gamma (\mu\to e\gamma)={\alpha\over 4}m_{\mu}^3\l |F_2^{(a)}|^2+|F_2^{(b)}|^2\r\;,
\end{equation}
where
\begin{equation}
\label{F2}
F_2^{(a)}={\alpha_1\over 4\pi M_{Bino}}\mu
m_{\mu}\tan{\beta}V_{23}V_{13}\left[G_2\l\tilde{m}^2_{{\sm L}},
\tilde{m}^2_{{\sm R}}-{\Lambda^2\over
8\pi^2}f_3(x)\Delta^2\r-G_2\l\tilde{m}^2_{{\sm L}},
\tilde{m}^2_{{\sm R}}\r\right]\;,
\end{equation}
\begin{equation}
\label{F2b}
F_2^{(b)}={\alpha_1\over 4\pi }      
m_{\mu}V_{23}V_{13}\left[G_1\l\tilde{m}^2_{{\sm R}}-{\Lambda^2\over
8\pi^2}f_3(x)\Delta^2\r-G_1\l\tilde{m}^2_{{\sm R}}\r\right]
\end{equation}
with 
$$
G_2(m_1^2,m_2^2)={g_2\l{m_1^2\over M_{Bino}^2}\r-g_2\l{m_2^2\over M_{Bino}^2}\r\over 
m_1^2-m_2^2}\;,\;\;g_2(r)={1\over 2(r-1)^3}[r^2-1-2r\ln{r}]\;,
$$ 
$$
G_1(m^2)=\frac{1}{M_{Bino}^2}g_1\l\frac{m^2}{M_{Bino}^2}\r\;,\;\;
g_1(r)=-\frac{1}{6(r-1)^4}[2+3r-6r^2+r^3+6r\ln{r}]\;.
$$

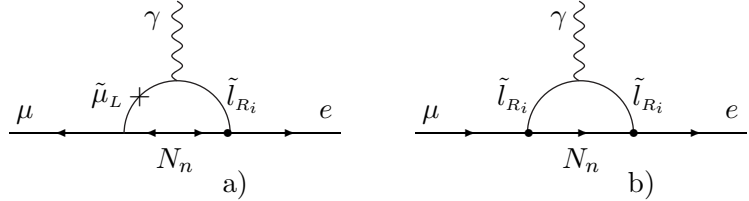
\begin{figure}[htb]
\unitlength=1mm
\linethickness{0.4pt}
\special{em:linewidth 0.4pt}
\begin{picture}(190.00,20.00)(-20,120)
\put(10.00,125.00){\line(1,0){44.00}}
\put(17.00,125.00){\vector(-1,0){1.00}}
\put(47.00,125.00){\vector(1,0){1.00}}
\CArc(91.50,354.00)(20.0,0,180)
\Photon(91.50,374)(91.50,404){2}{4}
\Photon(243.00,374)(243.00,404){2}{4}
\put(12.00,127.00){\makebox(0,0)[cb]{$\mu$}}
\put(52.00,127.00){\makebox(0,0)[cb]{$e$}}
\put(32.00,123.00){\makebox(0,0)[ct]{$N_n$}}
\put(29.00,140.00){\makebox(0,0)[cc]{$\gamma$}}
\put(23.00,130.00){\makebox(0,0)[cc]{$\tilde{\mu} _{\sm L}$}}
\put(41.00,130.00){\makebox(0,0)[cc]{$\tilde{l}_{{\sm R}_i}$}}
\put(64.00,125.00){\line(1,0){44.00}}
\put(71.00,125.00){\vector(1,0){1.00}}
\put(101.00,125.00){\vector(1,0){1.00}}
\CArc(243.50,354.00)(20.0,0,180)
\put(66.00,127.00){\makebox(0,0)[cb]{$\mu$}}
\put(106.00,127.00){\makebox(0,0)[cb]{$e$}}
\put(86.00,123.00){\makebox(0,0)[ct]{$N_n$}}
\put(94.00,120.00){\makebox(0,0)[ct]{b)}}
\put(40.00,120.00){\makebox(0,0)[ct]{a)}}
\put(83.00,140.00){\makebox(0,0)[cc]{$\gamma$}}
\put(77.00,130.00){\makebox(0,0)[cc]{$\tilde{l}_{{\sm R}_i}$}}
\put(95.00,130.00){\makebox(0,0)[cc]{$\tilde{l}_{{\sm R}_i}$}}
%\put(32.00,132.00){\circle*{1.00}}
%\put(25.00,125.00){\circle*{1.00}}
\put(39.00,125.00){\circle*{1.00}}
\put(79.00,125.00){\circle*{1.00}}
%\put(86.00,132.00){\circle*{1.00}}
\put(93.00,125.00){\circle*{1.00}}
\put(27.30,129.80){\makebox(0,0)[cc]{$+$}}
%\put(27.30,129.80){\circle*{1.00}}
%\put(91.00,129.80){\makebox(0,0)[cc]{$+$}}
\put(86.00,125.00){\vector(1,0){1.00}}
\put(29.00,125.00){\vector(-1,0){1.00}}
\put(35.00,125.00){\vector(1,0){1.00}}
\end{picture}
\caption{Diagrams contributing to  $\mu \to e\gamma$ decay}
\label{popa}
\end{figure}

At large $\tan{\beta}$ (say, $\tan{\beta}\sim 50$), the leading contribution
to $\mu\to e\gamma$ decay is given by the term~(\ref{F2})
which comes from the diagram
shown in Fig.~\ref{popa}a with left-right slepton mixing insertion. 
(We will comment on the validity of this approximation later on.) 
This contribution is enhanced by a factor of order 
${M_{Bino} \mu\tan{\beta} \over
\tilde{m}_{\sm L}^2}\sim 30$ in comparison with the term~(\ref{F2b}), 
coming from the diagram of Fig.~\ref{popa}b 
without chirality flip. 
In Fig.~\ref{popa}, $N_n$ denotes combinations of bino and higgsino.
In fact, the dominant effect comes from bino, as higgsino is 
significantly heavier in MGMM. 

The elements of the rotating matrix $V_{ij}$ can be found explicitly in our case.
The relevant matrix elements are 
$$
V_{i3}={Y^{(5)}_{4i}\over \Delta}\;.
$$
Hence, the functions~(\ref{F2}), (\ref{F2b}) can be written in the following form,
\begin{eqnarray}
\label{F2D}
F_2^{(a)}={\alpha_1\over 4\pi M_{Bino}}\mu m_{\mu}\tan{\beta}{Y^{(5)}_{41}Y^{(5)}_{42}
\over \Delta^2}\left[G_2\l\tilde{m}^2_{{\sm L}},
\tilde{m}^2_{{\sm R}}-{\Lambda^2\over
8\pi^2}f_3(x)\Delta^2\r-G_2\l\tilde{m}^2_{{\sm L}},
\tilde{m}^2_{{\sm R}}\r\right]\approx\\ {\alpha_1\over 4\pi }\mu m_{\mu}
\tan{\beta}\delta m_{12}^2
{\partial G_2\l\tilde{m}^2_{{\sm L}},
\tilde{m}^2_{{\sm R}}\r\over\partial\tilde{m}^2_{{\sm R}}}\;, \nonumber
\end{eqnarray}
\begin{eqnarray}
\label{F2Db}
F_2^{(b)}={\alpha_1\over 4\pi } m_{\mu}{Y^{(5)}_{41}Y^{(5)}_{42}
\over \Delta^2}\left[G_1\l\tilde{m}^2_{{\sm R}}-{\Lambda^2\over
8\pi^2}f_3(x)\Delta^2\r-G_1\l
\tilde{m}^2_{{\sm R}}\r\right]\approx\\ {\alpha_1\over 4\pi M_{Bino}} m_{\mu}
\delta m_{12}^2
{\partial G_1\l\tilde{m}^2_{{\sm R}}\r\over\partial\tilde{m}^2_{{\sm R}}}\;, \nonumber
\end{eqnarray}
where the approximate equalities hold at small $\Delta$ and correspond
to mass insertion approximation with respect to mixing of right
sleptons.  Equations (\ref{F2D}) and (\ref{F2Db}) and experimental
bounds~\cite{Particle_Data_Group} on the rate $\Gamma(\mu\to e\gamma)$
give limits on the parameters $Y^{(5)}_{41}$, $Y^{(5)}_{42}$ and
$Y^{(5)}_{43}$ which appear in the combinations
$Y^{(5)}_{41}Y^{(5)}_{42}$ and $\Delta$. The maximal allowed value of
the product $Y^{(5)}_{41}Y^{(5)}_{42}$ corresponds to the regime
$Y^{(5)}_{41}=Y^{(5)}_{42}$ and $Y^{(5)}_{43}=0$. In this regime
$\Delta^2=2Y^{(5)}_{41}Y^{(5)}_{42}$; we will see that $\Delta$ is
small there, so that approximate
equalities in eqs.~(\ref{F2D}) and (\ref{F2Db}) indeed hold. 
This means that conservative upper limit on the product
$Y^{(5)}_{41}Y^{(5)}_{42}$ can be obtained by making use 
of the mass insertion with respect to mixing of the 
right sleptons from the very beginning.
 The corresponding limits on the product
$\sqrt{Y^{(5)}_{41}Y^{(5)}_{42}f_3(x)}\sim
\sqrt{Y^{(5)}_{41}Y^{(5)}_{42}}x$ at $x$ not very close to one is
shown in Table~\ref{lepres}. One can see
that constraints on the mixing terms in the case of high $\tan{\beta}$ are
stronger. 

At  $Y^{(5)}_{41}\neq Y^{(5)}_{42}$ and/or $Y^{(5)}_{43}\neq 0$, one
has  $\Delta^2>2Y^{(5)}_{41}Y^{(5)}_{42}$ and it follows from the
explicit forms of $G_1$ and $G_2$ that
limits on the
product $\sqrt{Y^{(5)}_{41}Y^{(5)}_{42}}x$ are stronger than those
presented in Table~\ref{lepres}. The deviation from the mass insertion
results (approximate equalities in eqs.~(\ref{F2D}) and (\ref{F2Db}))
is not very strong, however, so Table~\ref{lepres} still gives an 
idea of the allowed values of Yukawa couplings.
\begin{table}[t]
\begin{center}
\begin{tabular}{|r|r|r|}
\hline
&$\tan{\beta}\sim 50$  & $\tan{\beta}\sim 1$ \\
\hline
${\bf 5+\overline{5}:}$&$\left |Y^{(5)}_{41}
Y^{(5)}_{42}\right |^{\vphantom{\big|}1/2}x<6\cdot
10^{-4}$
&$\left |Y^{(5)}_{41}Y^{(5)}_{42}\right |^{1/2}x<0.003$
\\[8pt]
\hline
${\bf 10+\overline{10}:}$&$\left 
|Y^{(10)}_{41}Y^{(10)}_{42}\right |^{\vphantom{\big|}1/2}x<0.002$
&$\left |Y^{(10)}_{41}Y^{(10)}_{42}\right |^{1/2}x<0.03$
\\[8pt]
\hline
\end{tabular}
\caption{The constraints on the Yukawa couplings coming from the
$\mu\to e\gamma$ decay at $\Lambda=100$~TeV for fundamental and 
 $\Lambda=50$~TeV for antisymmetric messengers in the cases of high and low
$\tan{\beta}$ and $x\lesssim 0.8$.}
\label{lepres}
\end{center}
\end{table}

The slepton mixing in the gauge-mediated models gives rise also to
$\mu\to e$ 
conversion.  The dominant contribution to its rate $\Gamma(\mu \to e)$
is given by penguin-type diagrams, while box diagrams are
suppressed by squark masses.  So, there is a simple relation
between the rates of $\mu\to e$ conversion and $\mu \to e\gamma$~\cite{we},
$$
\frac{\Gamma(\mu \to e)}{\Gamma(\mu \to e\gamma)}=
2.8\cdot10^{-2}\;,
$$
while the ratio of experimental limits \cite{Particle_Data_Group} is
$$
\frac{\Gamma(\mu \to e)^{lim}_{exp}}
{\Gamma(\mu \to e\gamma)^{lim}_{exp}}=1.1\cdot10^{-1}\;.
$$
Hence, the existing limit on $\mu - e$ - conversion gives weaker
 (by a factor of $1.4$) bound on the product of Yukawa couplings
 $|Y^{(5)}_{41}Y^{(5)}_{42}|^{1/2}$. 

Analogously, one can find limits coming from flavor violating $\tau$
decays. One finds, that for these decays the limits in the regimes when one of
the Yukawas is equal to zero and $\tan{\beta}\sim 50$ 
are also strong enough for the mass insertion
approximation with respect to $\Delta^2$ to be valid. The
corresponding limits on $Yx$ are shown
in Table~\ref{lepres1}. 
\begin{table}[htb]
\begin{center}
\begin{tabular}{|r|r|r|}
\hline
&$\tau\to \mu\gamma$ & $\tau\to e\gamma$ \\
\hline
${\bf 5+\overline{5}:}$&$\left |Y^{(5)}_{42}Y^{(5)}_{43}\right |^{1/2}x<0.01$
&$\left |Y^{(5)}_{41}Y^{(5)}_{43}\right |^{1/2}x<0.01$ \\[8pt]
\hline
${\bf 10+\overline{10}:}$&$\left |Y^{(10)}_{42}Y^{(10)}_{43}\right
|^{1/2}x
<0.04$
&$\left |Y^{(10)}_{41}Y^{(10)}_{43}\right |^{1/2}x<0.04$  \\[8pt]
\hline
\end{tabular}
\caption{The constraints on the Yukawa couplings coming from flavor
violating $\tau$ decays at $\Lambda=100$~TeV for fundamental and 
 $\Lambda=50$~TeV for antisymmetric messengers, $x\lesssim 0.8$ and
$\tan{\beta}=50$.}
\label{lepres1}
\end{center}
\end{table}
Another situation occurs at low $\tan{\beta}$. 
In the case of flavor violating $\tau$ decays large values of $\Delta$
even at $Y^{(5)}_{41}=0$ (or $Y^{(5)}_{42}=0$ depending on
the type of $\tau$ decay) are not restricted
experimentally and mass insertion technique is not valid. 
The upper bounds on the relevant combination of Yukawa couplings 
derived from
the experimental limits on $\Gamma (\tau\to \mu\gamma)$ and $\Gamma
(\tau\to e\gamma)$~\cite{Particle_Data_Group} are weaker 
%in the case of fundamental messenger 
%representation are shown in Fig.~\ref{shvah}. 
%\begin{figure}[b]
%\begin{center}
%{\psfig{file=pic2new.eps,width=12cm}}
%\end{center}
%\caption
%{The upper limits on Yukawa couplings from flavor changing
%$\tau$ - decays.  The dashed line is the limit on 
%$|Y^{(5)}_{41}Y^{(5)}_{43}|^{1/2}$ from $\tau \to e\gamma$ decay, the same
%line represents also the
%limit on $|Y^{(5)}_{42}Y^{(5)}_{43}|^{1/2}$ from $\tau \to \mu\gamma$ decay.
%The solid line corresponds to the theoretical constraint, 
%which is the 
%same for $|Y^{(5)}_{41}Y^{(5)}_{43}|^{1/2}$ and
% $|Y^{(5)}_{42}Y^{(5)}_{43}|^{1/2}$.}
%\label{shvah}
%\end{figure}
%From figure~\ref{shvah} one observes that these bounds are 
than theoretical constraints inherent in this model (see section~4.1).
At $\tan{\beta}\sim 1$ the experimental limits on the rates of flavor
changing $\tau$ decays are at least by order magnitude larger than
maximum rates allowed in this model. It is worth noting that this
value of suppression factor corresponds to the extreme limit
$Yx\gtrsim 1$. This factor scales as $(Yx)^{-4}$ at smaller value of
mixing. Hence, self-consistence of this model at low $\tan{\beta}$
requires that the rates $\Gamma(\tau
\to e\gamma)$ and $\Gamma(\tau \to \mu\gamma)$ are lower than the
present experimental limits.

There is another type of flavor violating processes, namely, 
oscillations of charged sleptons and sneutrinos. 
Pair production of sleptons (which decay into
leptons and bino) in $e^+-e^-$ annihilation at the Next Linear
Collider will result in acoplanar $l_i^+-l_i^-$ 
events with missing energy.  Recall, that the NLSP in the model with
low $\tan{\beta}$ is neutralino $\tilde{N}$, while in model with high
$\tan{\beta}$ the NLSP is $\tilde{\tau}$, so there 
will be additional $\tau$-leptons 
coming from bino decays in the latter case. In the presence of slepton
mixing, the slepton oscillations leading to lepton flavor
violating $l_i^{\pm}-l_j^{\mp}$ events, are possible
\cite{krasnikov}.  

For example, let us consider the case of fundamental messengers, low
$\tan{\beta}$ and 
$Y^{(5)}_{43}=0$. 
Oscillations of $\tilde{\mu}_{\sm R}$ and $\tilde{e}_{\sm R}$ 
are characterized by the corresponding mixing angle, which in this
model can be found from eq.~(\ref{delta-mixing-1}), 
\begin{equation}
\label{oscphi}
\tan{2\phi}=2\frac{|Y^{(5)}_{41}Y^{(5)}_{42}|}{|Y^{(5)}_{41}|^2-|Y^{(5)}_{42}|^2}
\end{equation}
If $Y^{(5)}_{41}\sim Y^{(5)}_{42}$ then the slepton mixing 
is close to maximal.  The
cross section of \linebreak $e^+e^- \to e^{\pm} \mu^{\mp}+2\tilde{N}$ may,
however, be suppressed even at large mixing if the lifetimes of
$\tilde{\mu}_{{\sm R}}$ and $\tilde{e}_{{\sm R}}$ are small compared to the period
of oscillations. The absence of such 
a suppression requires \cite{krasnikov}
\begin{equation}
\label{kras3}
2\Gamma_{sl} M_{e_{\sm R}} < |M^2_{e_{\sm R}}-M^2_{\mu_{\sm R}}|
\end{equation}
where $M_{e_{\sm R}}$ and $M_{\mu_{\sm R}}$ denote the true
slepton masses.  For slepton decay width $\Gamma_{sl}$ we have
\begin{equation}
\label{kras4}
\Gamma_{sl}=\frac{\alpha_1}{2}
\tilde{m}_{_{\sm R}}\bigg(1-\frac{M^2_{bino}}
{\tilde{m}^2_{_{\sm R}}}\bigg)^2\;,
\end{equation}
where $\tilde{m}_{_{\sm R}}$ is an average slepton mass. 
By making use of eqs.~(\ref{gaugin}), (\ref{scalmas}) and 
(\ref{delta-mixing-1}), 
it is straightforward to 
translate the condition (\ref{kras3}) 
into a condition imposed on Yukawa 
couplings $Y^{(5)}_{41}$ and $Y^{(5)}_{42}$.
For example, at small $x$ one has
\begin{equation}
\label{kras5}
(|Y^{(5)}_{41}|^2+|Y_{42}^{(5)}|^2)x^2 > 5\cdot10^{-6}
\end{equation}
This relation and limits shown in Table~\ref{lepres} imply that at low
$\tan{\beta}$ there is a fairly wide range of parameters in which $\mu
\to e\gamma$ decay and slepton oscillations are both allowed.  Note,
that unlike $\mu \to e\gamma$ rate, slepton oscillation parameters
$\sin 2\phi$ and $\frac{2\Gamma_{sl} M_{e_{{\sm R}}}}{|M_{e_{{\sm
R}}}^2-M_{\mu_{{\sm R}}}|^2}$ are independent of
$\Lambda$. Analogously, oscillations of $\tilde{\tau}$ into other
sleptons are also possible in the regime $Y^{(5)}_{43}\neq0$. In model
with high $\tan{\beta}$ the constraints on Yukawa couplings are
significantly stronger and slepton oscillations are suppressed in
comparison with the case of low $\tan{\beta}$.

%The decay $e^+e^- \to e^{\pm} \mu^{\mp}+4\tau$ would characterize
%slepton oscillations in the model with high $\tan{\beta}$. The estimates
%similar to the presented above show that with account of constraints
%on Yukawa couplings coming from lepton flavor violating decays the
%slepton oscillations is strongly suppressed in this model.
% At the Yukawa couplings of the same order the
%suppression occurs because of small lifetimes of sleptons, while at
%the couplings of different orders the rates of oscillation is
%suppressed by small mixing angle $\phi$ (see,~(\ref{oscphi})).

In the case of  antisymmetric messengers, the analysis is similar to
one presented above. 
The only difference is that flavor violation 
originates from the sector of left sleptons. 
In this case mixing in vertices with wino and zino arises
in addition to mixing in vertices with bino discussed above. Corresponding
amplitudes are smaller than bino mediated ones, because superpartners of
weak bosons are significantly heavier than bino in MGMM~\footnote{This
suppression survives even after taking into account that
$\alpha_2>\alpha_1$, because gaugino masses are proportional to
corresponding
$\alpha_i.$}. Since our purpose is to estimate 
the maximal allowed values of the mixing parameters it is
sufficient to take into account only bino mediated diagrams.

We present our results in Tables~\ref{lepres}, \ref{lepres1}, 
where the limits on the Yukawa 
couplings at $x\lesssim 0.8$ and $\Lambda=100$~TeV for fundamental
messengers and $\Lambda=50$~TeV for antisymmetric ones 
are given\footnote{We take smaller value of $\Lambda$ for
antisymmetric messengers, because superpartners are heavier in this
case at the same value of $\Lambda$.}. The experimental limits on the rates of the rare 
processes~\cite{Particle_Data_Group} are given for convenience 
in Table~\ref{rare} of Appendix {\bf C}. Let us make a few remarks
that apply to both cases of fundamental and antisymmetric messengers. 
As one can see from eq.~(\ref{F2}),
these bounds increase approximately linearly with $\Lambda$. 
Contributions to amplitudes coming from diagrams with chirality flip
are proportional to $\tan{\beta}$ and dominate at high
$\tan{\beta}$. Therefore corresponding bounds are inversely
proportional to $\sqrt{\tan{\beta}}$ at large $\tan{\beta}$. 
In similarity to the case of
fundamental messengers, for antisymmetric ones the rates of 
flavor changing $\tau$ decays are at the level of current experiments
in case of high $\tan{\beta}$ and are forbidden at low $\tan{\beta}$
regime. At 
$\tan{\beta}\sim 1$ the maximum rates of
flavor changing $\tau$ decays in model with antisymmetric messengers
are at least by order of magnitude smaller than the current
experimental limits.  

Finally, it is worth noting that there are corrections to 
eq.~(\ref{F2}) due to the diagrams with larger number of left-right mass
insertions. These corrections are proportional to $\tan{\beta}$ and
can be significant in high $\tan{\beta}$ region. 
To the leading order in $m_{\tau}/m_e$,
$m_{\tau}/m_{\mu}$ these corrections are proportional to
$V_{33}={Y^{(5)}_{43}\over\Delta}$. Consequently, in the case of
$\mu\to e\gamma$ decay they are not essential at 
$Y^{(5)}_{43}=0$ and the results given in Table~\ref{lepres} are not modified. In the case of flavor
changing $\tau$ decays these corrections make the limits, presented in
Table~\ref{lepres1} slightly stronger. However, numerical analysis shows
that the limits get modified very modestly even at large splitting, when very
light slepton appears. For example, for fundamental messengers 
at $\Lambda = 100$~TeV (which corresponds
to $\tilde{m}_{{\sm R}1}\sim 150$~GeV; we use the spectrum of MGMM
found in Ref.~\cite{borzumati}) the exact amplitude is 
only 1.4 times larger than one obtained by making use of eq.~(\ref{muegrate})
when the lightest slepton mass reaches its present experimental limit
of 60~GeV. 

\subsection{Quark sector}

We have found the bounds on various products 
of Yukawa couplings of messengers with quarks coming from the
requirement that corresponding mixing is consistent with the
present experimental limits
on the rare processes. We take into account only gluino mediated 
contributions to the rare processes. 
It is worth noting that there are also chargino and photino 
contributions to flavor changing processes. 
These contributions will make the limits presented below 
stronger by a factor of order one and they are not significant as far
as semi-quantitative estimates of the allowed values of Yukawa
couplings $Y_i$ are concerned. 
The experimental limits~\cite{Particle_Data_Group} are summarized 
in Table~\ref{rare} 
of Appendix~{\bf C}. 

To calculate these bounds we make use of the results of
Ref.~\cite{Masiero}, where flavor violation in the MSSM with 
general mass matrix was studied in the mass insertion approximation in
$\delta m_{ij}^2$. As we have seen in the
previous section, mass insertion approximation with respect to
chirality conserving terms works well up to 
very large mass splitting; in the latter regime exact calculations 
typically lead to slightly more stringent limits. 

However, it is worth noting that at high $\tan{\beta}$, in analogy to
the flavor changing lepton decays, it may be insufficient 
to take into account only one flavor
changing insertion. The point is that additional
left-right mixing insertion can significantly enhance some of the
amplitudes. This is not the case for box diagrams. Only this type of
diagrams contributes to amplitudes with $\Delta F=2$ and, consequently,
one can directly apply the results of Ref.~\cite{Masiero} for
$\Delta F=2$ processes. The limits on the Yukawa couplings coming from 
$\Delta F=2$ processes are basically independent of $\tan{\beta}$.  

In the case of $\Delta F=1$ processes ($b\to s\gamma$ decay and
CP-violation in $K^0$ decays) penguin diagrams give contributions as
well, and additional left-right mixing mass insertion is essential. Following
Ref.~\cite{Masiero}, we take it into account by introducing an
effective left-right mixing flavor changing insertion
$$
(\delta\tilde{m}_{ij}^2)_{eff}=\delta\tilde{m}_{ij}^2
\times{m_{q}\mu\tan{\beta}\over \tilde{m}_q^2}\;.
$$
Here $\delta\tilde{m}_{ij}^2$ is the original chirality conserving mass
insertion, $m_{q}$ and $\tilde{m}_{q}$ are  masses of the
corresponding quark and squark. Hence, $\Delta F=1$ processes depend
on the value of $\tan{\beta}$. 

We present our results in Tables~\ref{KB},\ref{Db} and \ref{bs}, 
where the limits on the Yukawa couplings 
for 
$x$ not very close to one ($x\lesssim 0.8$) and $\Lambda=100$~TeV are
shown. 
\begin{table}[htb]
$$
\begin{array}{|r|r|r|}
\hline
&
K^0-\bar{K}^0\;,\;\;\Delta F=2 
& 
K^0-\bar{K}^0,~~\epsilon\;,\;\;\Delta F=2\\
\hline
{\bf 5+\overline{5}:}
&
\left |\Re\l X^{(5)*}_{41}X^{(5)}_{42}\r ^2\right |^{\vphantom{\big|}1/4}x<0.08 
&
\left |\Im\l X^{(5)*}_{41}X^{(5)}_{42}\r^2\right 
|^{\vphantom{\big|}1/4}x<0.02\\[8pt]
\hline
{\bf 10+\overline{10}:}&
\left |\Re\l W^{(10)*}_{41}W^{(10)}_{42}\r^2\right
|^{\vphantom{\big|}1/4}x
<0.1
&
\left |\Im\l X^{(10)*}_{41}X^{(10)}_{42}\r^2\right 
|^{\vphantom{\big|}1/4}x<0.05\\[8pt]
\hline
{\bf 10+\overline{10}:}&
\left |\Re\l X^{(10)*}_{41}X^{(10)}_{42}\r^2\right
|^{\vphantom{\big|}1/4}x
<0.2 
&
\left |\Im\l W^{(10)*}_{41}W^{(10)}_{42}\r^2\right 
|^{\vphantom{\big|}1/4}x<0.04\\[8pt]
\hline
{\bf 10+\overline{10}:}&
\left |\Re X^{(10)*}_{41}X^{(10)}_{42}
W^{(10)*}_{41}W^{(10)}_{42}\right |^{\vphantom{\big|}1/4}x<0.04
&
\left |\Im\l X^{(10)*}_{41}X^{(10)}_{42}
W^{(10)*}_{41}W^{(10)}_{42}\r\right |^{\vphantom{\big|}1/4}x<0.01\\ [8pt]
\hline
\end{array}
$$
\caption{The constraints on Yukawa couplings coming from
$K^0-\bar{K}^0$ mixing at $\Lambda=100$~TeV for fundamental messengers
and at $\Lambda=50$~TeV for antisymmetric ones, $x\lesssim 0.8$.}
\label{KB}
\end{table}

\begin{table}[htb]
$$
\begin{array}{|r|r|r|}
\hline
&B^0-\bar{B}^0\;,\;\;\Delta F=2 
& 
D^0-\bar{D}^0\;,\;\;\Delta F=2\\
\hline
{\bf 5+\overline{5}:}&
\left |\Re\l X^{(5)*}_{41}X^{(5)}_{43}\r ^2\right |^{1/4}x<0.1
&
\left |\Re\l X^{(5)*}_{41}X^{(5)}_{42}\r ^2\right
|^{\vphantom{\big|}1/4}
x<0.1\\[8pt]
\hline
{\bf 10+\overline{10}:}&
\left |\Re\l X^{(10)*}_{41}X^{(10)}_{43}\r^2\right |^{1/4}x<0.3
&
\left |\Re\l X^{(10)*}_{41}X^{(10)}_{42}\r^2\right
|^{\vphantom{\big|}1/4}x
<0.3\\[8pt]

\hline
{\bf 10+\overline{10}:}&
\left |\Re\l W^{(10)*}_{41}W^{(10)}_{43}\r^2\right |^{1/4}x<0.2
&
\left |\Re\l X^{(10)*}_{14}X^{(10)}_{24}\r^2\right
|^{\vphantom{\big|}1/4}x
<0.2\;
\\[8pt]
\hline
{\bf 10+\overline{10}:}&
\left |\Re X^{(10)*}_{41}X^{(10)}_{43}
W^{(10)*}_{41}W^{(10)}_{43}\right |^{1/4}x<0.1
& 
\left |\Re X^{(10)*}_{41}X^{(10)}_{42}
X^{(10)*}_{14}X^{(10)}_{24}\right |^{\vphantom{\big|}1/4}x<0.1
\\[8pt]
\hline
\end{array}
$$
\caption{The constraints on Yukawa couplings coming from
$B^0-\bar{B}^0$ mixing and $D^0-\bar{D}^0$ mixing 
at the same values of parameters as in Table~\ref{KB}.}
\label{Db}
\end{table}

There are also limits on CP-violating terms (see Tables~\ref{KB},~\ref{CP})
coming from the analysis of $K^0-\bar{K}^0$ system and $K\to\pi\pi$
decays. 
\begin{table}[htb]
$$
\begin{array}{|r|r|r|}
\hline
&\epsilon'/\epsilon,\ \tan{\beta}\sim 50 &
\epsilon'/\epsilon,\ \tan{\beta}\sim 1 
\\
\hline
{\bf 5+\overline{5}:}& 
\left |\Im X^{(5)*}_{41}X^{(5)}_{42}\right |^{1/2}x<0.05
& 
\left |\Im X^{(5)*}_{41}X^{(5)}_{42}\right |^{1/2}x<0.3\\[8pt]
\hline
{\bf 10+\overline{10}:}& 
\left |\Im X^{(10)*}_{41}X^{(10)}_{42}\right |^{1/2}x<0.09
& 
\left |\Im X^{(10)*}_{41}X^{(10)}_{42}\right |^{1/2}x<0.9\\[8pt]
\hline
{\bf 10+\overline{10}:}& 
\left |\Im W^{(10)*}_{41}W^{(10)}_{42}\right |^{1/2}x<0.06
&
\left |\Im W^{(10)*}_{41}W^{(10)}_{42}\right |^{1/2}x<0.6
\\ [8pt]
\hline
\end{array}
$$
\caption{The constraints on Yukawa couplings coming from $\Delta F=1$
CP-violating processes at high and low $\tan{\beta}$. 
The parameters are the same as in
Table~\ref{KB}.}
\label{CP}
\end{table}
At high (low) $\tan{\beta}$ 
the limits from the latter process are stronger (weaker) than those coming
from $K^0-\bar{K}^0$ system at the level of current experiments. 

A typical constraint on Yukawa couplings in the quark sector is
$Yx\lesssim 0.1$. 
It is clear from Tables \ref{KB} -- \ref{bs} that different experiments are
sensitive, generally speaking, to different combinations of Yukawa 
couplings and CP-violating phases. However, one may notice that
 $K^0-\bar{K}^0$ system is presently a particularly good probe of the
messenger-matter mixing in the quark sector.

It turns out that the contribution to ${\mbox Br}(b\to s\gamma)$  due to the
messenger-matter mixing at low $\tan{\beta}$ 
is about $10^{-6}$ which is $10^2$ times smaller than
current experimental uncertainties in the region of parameters allowed by
theoretical bounds. 
Meanwhile there are contributions
of order $10^{-4}$
to ${\mbox Br}(b\to s\gamma)$
in gauge mediated models without mixing~(see, e.g., 
Refs.\cite{thomas,borzumati}). Correspondingly, messenger-matter mixing 
is not significant for this process in the region of low $\tan{\beta}$. 

Limits coming from $b\to s\gamma$ decay in the region of high $\tan{\beta}$
are shown in Table~\ref{bs}.

\begin{table}[htb]
$$
\begin{array}{|r|r|}
\hline
&
b\to s\gamma\;,\;\;\Delta F=1\\

\hline
{\bf 5+\overline{5}:}&
\left |X^{(5)*}_{42}X^{(5)}_{43}\right |^{1/2}x<0.2 \\[8pt]

\hline
{\bf 10+\overline{10}:}&
\left | X^{(10)*}_{42}X^{(10)}_{43}\right |^{1/2}x<0.3 \\[8pt]

\hline
{\bf 10+\overline{10}:}&
\left |\Re\l W^{(10)*}_{42}W^{(10)}_{43}\r^2\right |^{1/4}x<0.2\ \\[8pt]

\hline

\end{array}
$$
\caption{The constraints on Yukawa couplings coming from
$B^0-\bar{B}^0$ mixing and $b\to s\gamma$ decay at the same values of
parameters as in Table~\ref{KB}; $\tan{\beta}$=50.}
\label{bs}
\end{table}

Bounds on $Y$ coming from $\Delta F=2$ processes scale 
as $\sqrt{\Lambda}$. This scaling comes from the fact that
corresponding four-fermion operators in the effective Hamiltonian
originate from diagrams with two mixing insertions $\delta m^2_{ij}$
and the coefficients in front of these operators are proportional to 
$\frac{Y^4}{\Lambda^2}$.
Analogously, the limits coming from $\Delta F=1$ processes scale as
$\Lambda$. Bounds on $Y$ from $\Delta F=1$ processes at high
$\tan{\beta}$ are inversely proportional to $\sqrt{\tan{\beta}}$.

\section{Concluding remarks}
We have considered mixing between the usual matter and messengers
belonging to either the fundamental or antisymmetric complete SU(5)
multiplets in the Minimal Gauge Mediated Model. Limits
on the corresponding coupling constants coming from various flavor 
violating processes in lepton and quark sector and CP violating processes 
in 
quark sector have been found. 

These limits depend on the ratio $x$ of the supersymmetry breaking
parameter $\Lambda$
and messenger scale. The final results were
presented for relatively small ratio: $0<x\lesssim 0.8$. 
It is straightforward to
extend this  analysis to $x$ close to 1, and the results do not
change drastically. 

We have seen that small value of $\tan{\beta}$ naturally 
appears in MGMM 
with messenger-matter mixing. 
This fact 
may help to construct a realistic 
$SU(5)$ Grand 
Unified Theory, as models with 
low $\tan{\beta}$ (with incomplete messenger multiplets) are unifiable
with long proton lifetime~\cite{mes-split}, unlike models with 
large\footnote{In 
the case 
of complete messenger multiplets, low $\tan{\beta}$ does not save proton from 
fast decay in the 
framework of $SU(5)$ GUT~\cite{su5}} $\tan{\beta}$.

Another consequence
of small $\tan{\beta}$ is 
the reduction of the mixing between $\tilde{\tau}_{\sm R}$
and $\tilde{\tau}_{\sm L}$. As a result, 
now  the NLSP can be photino, 
rather than the right
slepton and, consequently, the 
predictions of
this model for 
 collider experiments can be significantly different. For example, 
this case is more suitable for explaining the CDF event~\cite{CDF}.

We have seen in this paper that in
the case of large enough  mixing in the MGMM with fundamental messengers,
there appears one and only one pair of light left squarks. 
This fact 
provides an interesting signature 
of MGMM with 
messenger-matter mixing. For $m_{light}<1$~TeV one 
has $\tan{\beta}\simeq 1$ in this model.
One expects that similar phenomenon exists also in models with more 
complicated messenger content.

There are two types of constraints on the allowed region in the space
of 
messenger-matter Yukawa couplings.
All experimental bounds coming from the flavor physics 
limit only 
the products of
different Yukawa couplings $|Y_iY_j|^{1/2}$ but not 
$Y_i$ separately. 
On the other hand, theoretical bounds, coming from the requirement of
positivity of the scalar masses, 
correspond to spherical regions  $\sum |Y_i|^2<$~const
in the space of Yukawa couplings. 

The rates of some rare processes  depend crucially 
on the value of $\tan{\beta}$. 
At high $\tan{\beta}\sim 50$ and $x\gtrsim 0.1$, 
experimentally accessible values of most of the
messenger-matter Yukawa couplings are in the interesting 
range $10^{-3}\div 10^{-1}$. A particularly sensitive probe of 
messenger-matter mixing is muon flavor violation. For example, in the
case of fundamental messengers the present limit
on $\mu\to e\gamma$ decay rate implies  
$|Y_{41}^{(5)}Y_{42}^{(5)}|x^2<4.0\times
10^{-7}$ while the bound from the $\mu\to e$ conversion is weaker by a factor 
of 2. Future experiments on $\mu\to e\gamma$ decay and $\mu\to e$ conversion
are quite promising from the point of view 
of MGMM with messenger-matter mixing.  

The case of low $\tan{\beta}$ is somewhat different. 
The experimentally accessible values of most of the
messenger-matter Yukawa couplings are one order of magnitude higher 
than in previous case, with the exception of $\Delta F=2$ 
processes whose sensitivity remains the same. The present limit
on $\mu\to e\gamma$ decay rate implies  
$|Y_{41}^{(5)}Y_{42}^{(5)}|x^2<10^{-5}$. There 
are basically no experimental constraints 
coming from $\tau\to e\gamma$, $\tau\to\mu\gamma$
decays and the corresponding mixing terms are limited by the
 self-consistency conditions inherent in the theory.
Even at the extreme values $Y_i x\sim 1$ branching ratios of the
lepton flavor violating $\tau$ decays 
must be order of magnitude smaller than 
existing
experimental bounds. Hence, at low $\tan{\beta}$ this model forbids 
flavor violating $\tau$ decays at 
the level of the next generation experiments.  
There are 
essentially no constrains coming from $b\to s\gamma$ 
as well, depending on the type of Yukawa couplings, 
the contributions of  messenger-matter mixing
 to the $b\to s\gamma$ rate are two orders of magnitude
smaller than experimental
 uncertainties in the case of low $\tan{\beta}$.

Finally, it is worth noting that the estimates of the allowed range of
the mixing parameters presented in this paper are expected to remain 
qualitatively the same in more general gauge mediated models than MGMM. 

\section{Acknowledgments}
The authors are indebted V.A.Rubakov for stimulating interest and 
helpful 
suggestions. We thank F.L.Bezrukov, M.L.Libanov, P.G.Tinyakov and
S.V.Troitsky for useful 
discussions. This work is supported in
part by Russian Foundation for Basic Research grant
96-02-17449a, by the  INTAS
     grant 96-0457 within the research program of the
     International Center for Fundamental Physics in
     Moscow and by ISSEP fellowships. 

\section{Appendix}
{\bf A ~~~ Fermion mass matrix} 

Here we present the explicit forms of fermion mass matrices in the
model with fundamental messengers. 
The lepton mass matrix that includes left and
right fermionic messengers has the following form
\begin{equation}
\label{fermi-d}
{\cal M}^l_{f} = {\cal U}^l_{f{\sm L}}{\cal D}^l_{f}{\cal U}^l_
{f{\sm R}}
\end{equation}
 where
\begin{equation}
\label{ufl}
{\cal U}^l_{f{\sm L}} = 
\left(
\begin{array}{cccc}
1& 0& 0& -\frac{y_{e}y_{1}^*}{\lambda^2S^2} \\[2pt]
0& 1& 0& -\frac{y_{\mu}y_{2}^*}{\lambda^2S^2} \\[2pt]
0& 0& 1& -\frac{y_{\tau}y_{3}^*}{\lambda^2S^2} \\[2pt]
\frac{y_{e}y_{1}}{\lambda^2S^2}&  
\frac{y_{\mu}y_{2}}{\lambda^2S^2}&  
\frac{y_{\tau}y_{3}}{\lambda^2S^2}& 1 
\end{array}
\right)
\end{equation}
and
\begin{equation}
\label{ufr}
{\cal U}^l_{f{\sm R}} = 
\left(
\begin{array}{cccc}
1 - \frac{|y_{1}|^2}{2\lambda^2S^2}& -
\frac{y_{1}^*y_{2}}{2\lambda^2S^2}& 
-\frac{y_{1}^*y_{3}}{2\lambda^2S^2}& -
\frac{y_{1}^*}{\lambda S}\\[2pt]
-\frac{y_{1}y_{2}^*}{2\lambda^2S^2}&
1 - \frac{|y_{2}|^2}{2\lambda^2S^2}&
-\frac{y_{2}^*y_{3}}{2\lambda^2S^2}& -
\frac{y_{2}^*}{\lambda S}\\[2pt]
-\frac{y_{1}y_{3}^*}{2\lambda^2S^2}& 
-\frac{y_{2}y_{3}^*}{2\lambda^2S^2}& 
1 - \frac{|y_{3}|^2}{2\lambda^2S^2}& -
\frac{y_{3}^*}{\lambda S}\\[2pt]
\frac{y_{1}}{\lambda S}& 
\frac{y_{2}}{\lambda S}&
\frac{y_{3}}{\lambda S}& 
1 - \frac{|y_{1}|^2 + |y_{2}|^2 + 
|y_{3}|^2}{2\lambda^2S^2}
\end{array}
\right)
\end{equation}
are mixing matrices to the leading order in $\frac{y}{\lambda
S}$. Here $v_{\sm U}$ and $v_{\sm D}$ are the Higgs expectation
values,
$$y_{i}=Y_{4i} v_{\sm{D}},~~ y_{e,\mu,\tau}=Y_{e,\mu,\tau} 
v_{\sm{D}}$$
and
\begin{equation}
\label{d}
\begin{array}{rcl}
{\cal D}^l_{f} & = & diag\Bigl(y_{e}\bigl(1 - \frac{y_{e}}
{\lambda S}\bigl| 
\frac{y_{1}}{\lambda S}\bigr|^2\bigr),y_{\mu}\bigl(1 -
 \frac{y_{\mu}}
{\lambda S}\bigl| 
\frac{y_{2}}{\lambda S}\bigr|^2\bigr),
y_{\tau}\bigl(1 - \frac{y_{\tau}}
{\lambda S}\bigl| 
\frac{y_{3}}{\lambda S}\bigr|^2\bigr),\\ & & 
\lambda S\bigl(1 + 
\bigl|\frac{y_{1}y_{e}}{\lambda^2S^2}\bigr|^2 + 
\bigl|\frac{y_{2}y_{\mu}}{\lambda^2S^2}\bigr|^2 +
\bigl|\frac{y_{3}y_{\tau}}{\lambda^2S^2}\bigr|^2\bigl)\Bigr)
\end{array}
\end{equation}
is the matrix of mass eigenvalues.

For down-quark-like fermions one has 
\begin{equation}
\label{fermi-d-q}
{\cal M}^q_{f} = {\cal U}^q_{f{\sm L}}{\cal D}^q_{f}{\cal U}^q_
{f{\sm R}},
\end{equation}
 where
\begin{equation}
\label{ufl-q}
{\cal U}^q_{f{\sm L}} = 
\left(
\begin{array}{cccc}
1& 0& 0& -\frac{y_{d}y_{1}^*}{\lambda^2S^2} \\[2pt]
0& 1& 0& -\frac{y_{s}y_{2}^*}{\lambda^2S^2} \\[2pt]
0& 0& 1& -\frac{y_{b}y_{3}^*}{\lambda^2S^2} \\[2pt]
\frac{y_{d}y_{1}}{\lambda^2S^2}&  
\frac{y_{s}y_{2}}{\lambda^2S^2}&  
\frac{y_{b}y_{3}}{\lambda^2S^2}& 1 
\end{array}
\right)
\end{equation}
and
\begin{equation}
\label{ufr-q}
{\cal U}^q_{f{\sm R}} = 
\left(
\begin{array}{cccc}
1 - \frac{|y_{1}|^2}{2\lambda^2S^2}& -
\frac{y_{1}^*y_{2}}{2\lambda^2S^2}& 
-\frac{y_{1}^*y_{3}}{2\lambda^2S^2}& -
\frac{y_{1}^*}{\lambda S}\\[2pt]
-\frac{y_{1}y_{2}^*}{2\lambda^2S^2}&
1 - \frac{|y_{2}|^2}{2\lambda^2S^2}&
-\frac{y_{2}^*y_{3}}{2\lambda^2S^2}& -
\frac{y_{2}^*}{\lambda S}\\[2pt]
-\frac{y_{1}y_{3}^*}{2\lambda^2S^2}& 
-\frac{y_{2}y_{3}^*}{2\lambda^2S^2}& 
1 - \frac{|y_{3}|^2}{2\lambda^2S^2}& -
\frac{y_{3}^*}{\lambda S}\\[2pt]
\frac{y_{1}}{\lambda S}& 
\frac{y_{2}}{\lambda S}&
\frac{y_{3}}{\lambda S}& 
1 - \frac{|y_{1}|^2 + |y_{2}|^2 + 
|y_{3}|^2}{2\lambda^2S^2}
\end{array}
\right)
\end{equation}
are mixing matrices to the leading order in $\frac{y}{\lambda
S}$. Here 
$$y_{i}=Y_{4i} v_{\sm{D}},~~ y_{d,s,b}=Y_{d,s,b} 
v_{\sm{D}}$$
\begin{equation}
\label{d-q}
\begin{array}{rcl}
{\cal D}^q_{f} & = & \makebox{diag}\Bigl(y_{d}\bigl(1 - \frac{y_{d}}
{\lambda S}\bigl| 
\frac{y_{1}}{\lambda S}\bigr|^2\bigr),y_{s}\bigl(1 -
 \frac{y_{s}}
{\lambda S}\bigl| 
\frac{y_{2}}{\lambda S}\bigr|^2\bigr),
y_{b}\bigl(1 - \frac{y_{b}}
{\lambda S}\bigl| 
\frac{y_{3}}{\lambda S}\bigr|^2\bigr),\\ & & 
\lambda S\bigl(1 + 
\bigl|\frac{y_{1}y_{d}}{\lambda^2S^2}\bigr|^2 + 
\bigl|\frac{y_{2}y_{s}}{\lambda^2S^2}\bigr|^2 +
\bigl|\frac{y_{3}y_{b}}{\lambda^2S^2}\bigr|^2\bigl)\Bigr).
\end{array}
\end{equation}
The mass matrices in the model with antisymmetric messengers have
similar structure.

{\bf B ~~~ Scalar mass matrix}

The tree level mass term of sleptons including the scalar messengers
with the quantum numbers of left leptons 
has the following form
$$
{\cal V}^l_{sc} = s^l{\cal M}^l_{sc}s^{l\dagger}
$$
where
\begin{equation}
\label{basis}
s^l = (\tilde{e}_{{\sm L}},\tilde{\mu}_{{\sm L}},\tilde{\tau}_
{{\sm L}},
\tilde{e}_{{\sm R}}^*,\tilde{\mu}_{{\sm R}}^*,\tilde{\tau}_
{{\sm R}}^*,l,\bar{l}^*)
\end{equation}
are the scalar fields (we use the same notation for left selectrons as
for full left doublets), and 
$$
{\cal M}^l_{sc}=\left(
\begin{array}{cccccccc}
\tilde{m}_{e_{L}}^2 & 0 & 0 & \mu Y_{e}\upsilon_{\smU}
 & 0 & 0 & y_{e}y_{1}^* & 0\\
0 & \tilde{m}_{\mu_{L}}^2 & 0 & 0 & \mu Y_{\mu}
\upsilon_{\smU} & 0 & 
y_{\tau}y_{2}^* & 0\\
0 & 0 & \tilde{m}_{\tau_{L}}^2 & 0 & 0 &
 \mu Y_{\tau}\upsilon_{\smU} & y_{\tau}y_{3}^* & 0\\
\mu Y_{e}\upsilon_{\smU} & 0 & 0 & \tilde{m}_
{e_{R}}^2 & y_{1}^*y_{2} & y_{1}^*y_{3} & 
\mu Y_{1}^*\upsilon_{\smU} & \lambda Sy_{1}^*\\
0 & \mu Y_{\mu}\upsilon_{\smU} & 0 & y_{1}y_{2}^* &
 \tilde{m}_{\mu_{R}}^2 &
 y_{2}^*y_{3} & 
\mu Y_{2}^*\upsilon_{\smU} & \lambda Sy_{2}^*\\
0 & 0 & \mu Y_{\tau}\upsilon_{\smU} & y_{1}y_{3}^* 
& y_{2}y_{3}^* &
\tilde{m}_{\tau_{R}}^2 & \mu Y_{3}^*\upsilon_{\smU} &
 \lambda Sy_{3}^*\\
y_{e}y_{1} & y_{\mu}y_{2} & y_{\tau}y_{3} & \mu Y_{1}
\upsilon_{\smU} & 
\mu Y_{2}\upsilon_{\smU} & \mu Y_{3}\upsilon_{\smU} & 
\lambda^2S^2 & - \lambda F\\
0 & 0 & 0 & \lambda Sy_{1} & \lambda Sy_{2} & \lambda 
Sy_{3} & 
- \lambda F & \lambda^2S^2
\end{array}
\right)
$$
 The mass matrix of scalar fields with quantum numbers of down quarks
may be written as follows 
$$
{\cal V}^q_{sc} = s^q{\cal M}^q_{sc}s^{q\dagger},
$$
where 
\begin{equation}
\label{basis-q}
s = 
(\tilde{d}_{{\sm R}}^*,\tilde{s}_{{\sm R}}^*,\tilde{b}_{{\sm R}}^*,
\tilde{d}_{{\sm L}},\tilde{s}_{{\sm L}},\tilde{b}_{{\sm L}},
q,\bar{q}^*)
\end{equation}
are scalar fields, and 
$$
{\cal M}_{sc}=\left(
\begin{array}{cccccccc}
\tilde{m}_{d_{L}}^2 & 0 & 0 & \mu Y_{d}\upsilon_{\smU}
 & 0 & 0 & y_{d}y_{1}^* & 0\\
0 & \tilde{m}_{s_{L}}^2 & 0 & 0 & \mu Y_{s}
\upsilon_{\smU} & 0 & 
y_{b}y_{2}^* & 0\\
0 & 0 & \tilde{m}_{b_{L}}^2 & 0 & 0 &
 \mu Y_{b}\upsilon_{\smU} & y_{b}y_{3}^* & 0\\
\mu Y_{d}\upsilon_{\smU} & 0 & 0 & \tilde{m}_
{d_{R}}^2 & y_{1}^*y_{2} & y_{1}^*y_{3} & 
\mu Y_{1}^*\upsilon_{\smU} & \lambda Sy_{1}^*\\
0 & \mu Y_{s}\upsilon_{\smU} & 0 & y_{1}y_{2}^* &
 \tilde{m}_{s_{R}}^2 &
 y_{2}^*y_{3} & 
\mu Y_{2}^*\upsilon_{\smU} & \lambda Sy_{2}^*\\
0 & 0 & \mu Y_{b}\upsilon_{\smU} & y_{1}y_{3}^* 
& y_{2}y_{3}^* &
\tilde{m}_{b_{R}}^2 & \mu Y_{3}^*\upsilon_{\smU} &
 \lambda Sy_{3}^*\\
y_{d}y_{1} & y_{s}y_{2} & y_{b}y_{3} & \mu Y_{1}
\upsilon_{\smU} & 
\mu Y_{2}\upsilon_{\smU} & \mu Y_{3}\upsilon_{\smU} & 
\lambda^2S^2 & - \lambda F\\
0 & 0 & 0 & \lambda Sy_{1} & \lambda Sy_{2} & \lambda 
Sy_{3} & 
- \lambda F & \lambda^2S^2
\end{array}
\right)
$$

{\bf C ~~~ Experimental data}

Experimental results used in this paper are summarized in
Table~\ref{rare}. 
\begin{table}[htb]
$$
\begin{array}{|r|r|}
\hline
K^0-\bar{K}^0 & \Delta m_{\sm K}=(3.491\pm 0.009)\times 10^{-12}~MeV
\\
\hline 
D^0-\bar{D}^0 & \Delta m_{\sm D}<1.38\times10^{-10}~MeV \\
\hline
 B^0-\bar{B}^0 & \Delta m_{\sm B}=(3.12\pm 0.21)\times 10^{-10}~MeV \\
\hline
\mu\to e\gamma & \mbox{Br}(\mu\to e\gamma)< 4.9\times 10^{-11}\\
\hline
\tau\to e\gamma & \mbox{Br}(\tau\to e\gamma)<2.7\times 10^{-6} \\
\hline
\tau\to \mu\gamma & \mbox{Br}(\tau\to \mu\gamma)<3.0\times 10^{-6} \\
\hline
b\to s\gamma & \mbox{Br}(b\to s\gamma)=(4.2\pm1.0)\times 10^{-4} \\
\hline
\mu\to e\; conversion & \Gamma(\mu^-Ti\to e^-Ti)/\Gamma(\mu^-Ti\to
all)<4\times 10^{-12}\\
\hline
CP-violation & \epsilon=(2.268\pm 0.019)\times 10^{-3} \\
\hline
CP-violation & \epsilon'/\epsilon=(1.5\pm 0.8)\times 10^{-3} \\
\hline
\end{array}
$$
\caption{The experimental data on the flavor and CP-violating
processes.}
\label{rare}
\end{table}
%%%%%%%%%%%%%%%%%%%%%%%%%%%%%%%%%%%%%%%%%%%%%%%%%%%%%% 
\def\ijmp#1#2#3{{\it Int. Jour. Mod. Phys. }{\bf #1~}(19#2)~#3}
\def\pl#1#2#3{{\it Phys. Lett. }{\bf B#1~}(19#2)~#3}
\def\zp#1#2#3{{\it Z. Phys. }{\bf C#1~}(19#2)~#3}
\def\prl#1#2#3{{\it Phys. Rev. Lett. }{\bf #1~}(19#2)~#3}
\def\rmp#1#2#3{{\it Rev. Mod. Phys. }{\bf #1~}(19#2)~#3}
\def\prep#1#2#3{{\it Phys. Rep. }{\bf #1~}(19#2)~#3}
\def\pr#1#2#3{{\it Phys. Rev. }{\bf D#1~}(19#2)~#3}
\def\np#1#2#3{{\it Nucl. Phys. }{\bf B#1~}(19#2)~#3}
\def\mpl#1#2#3{{\it Mod. Phys. Lett. }{\bf #1~}(19#2)~#3}
\def\arnps#1#2#3{{\it Annu. Rev. Nucl. Part. Sci. }{\bf #1~}(19#2)~#3}
\def\sjnp#1#2#3{{\it Sov. J. Nucl. Phys. }{\bf #1~}(19#2)~#3}
\def\jetp#1#2#3{{\it JETP Lett. }{\bf #1~}(19#2)~#3}
\def\app#1#2#3{{\it Acta Phys. Polon. }{\bf #1~}(19#2)~#3}
\def\rnc#1#2#3{{\it Riv. Nuovo Cim. }{\bf #1~}(19#2)~#3}
\def\ap#1#2#3{{\it Ann. Phys. }{\bf #1~}(19#2)~#3}
\def\ptp#1#2#3{{\it Prog. Theor. Phys. }{\bf #1~}(19#2)~#3}
\def\spu#1#2#3{{\it Sov. Phys. Usp.}{\bf #1~}(19#2)~#3}
\def\epj#1#2#3{{\it Eur.\ Phys.\ J. }{\bf C#1~} #3 (19#2)}
%%%%%%%%%%%%%%%%%%%%%%%%%%%%%%%%%%%%%%%%%%%%%%%%%%%%%%%

\end{document}